\documentclass[fleqn,usenatbib]{mnras}

\usepackage{newtxtext,newtxmath}

\usepackage[T1]{fontenc}
\usepackage{ae,aecompl}

\usepackage{graphicx}	
\usepackage{amsmath}	
\usepackage{amssymb}	
\usepackage{array}
\usepackage{booktabs}
\usepackage{enumitem}
\usepackage{hyperref}
\usepackage{bigstrut}
\usepackage{threeparttable}
\usepackage{multirow}
\usepackage{xcolor}
\usepackage{textcomp}

\newcommand{\nnone}{$n_0$ = 100$\,\mathrm{cm^{-3}}$}
\newcommand{\nfull}{$n_0$ = $2\times10^{14}\,\mathrm{cm^{-3}}$}
\newcommand{\nzero}{$n_0$ = $1\times10^{12}\,\mathrm{cm^{-3}}$}
\newcommand{\lime}{{\sc lime}}
\newcommand{\radmc}{{\sc radmc-3d}}
\newcommand{\casa}{{\sc casa}}
\newcommand{\alma}{ALMA}
\newcolumntype{C}[1]{>{\centering\let\newline\\\arraybackslash\hspace{0pt}}m{#1}}

\defcitealias{Evans&Ilee2015}{Paper I}

\title[GIs in a protosolar-like disc II]{Gravitational instabilities in a protosolar-like disc II: continuum emission and mass estimates}

\author[M. G. Evans et al.]{
M.\ G.\ Evans$^{1}$\thanks{E-mail: py09mge@leeds.ac.uk},
J.\ D.\ Ilee$^{2}$, 
T.\ W.\ Hartquist$^{1}$, 
P.\ Caselli$^{4}$, 
L.\ Sz\H{u}cs$^{4}$, 
\newauthor
S.\ J.\ D.\ Purser$^{1}$
A.\ C.\ Boley$^{3}$,
R.\ H.\ Durisen$^{5}$ and
J.\ M.\ C.\ Rawlings$^{6}$
\\
$^{1}$School of Physics \& Astronomy, University of Leeds, Leeds LS2 9LN, UK\\
$^{2}$Institute of Astronomy, Madingley Road, Cambridge CB3 0HA, UK\\
$^{3}$Department of Physics \& Astronomy, 6224 Agricultural Road, Vancouver, BC V6T 1Z1, Canada\\
$^{4}$Max-Planck-Institut f$\ddot{u}$r extraterrestrische Physik, Giessenbachstrasse, 85741 Garching bei M$\ddot{u}$nchen, Germany\\
$^{5}$Department of Astronomy, Indiana University, 727 East 3rd Street, Swain West 319, Bloomington, IN 47405, USA\\
$^{6}$Department of Physics \& Astronomy, University College London, London WC1E 6BT, UK
}

\date{Accepted 2017 May 30. Received 2017 May 25; in original form 2017 April 6}

\pubyear{2017}

\begin{document}
\label{firstpage}
\pagerange{\pageref{firstpage}--\pageref{lastpage}}
\maketitle

\begin{abstract}
Gravitational instabilities (GIs) are most likely a fundamental process during the early stages of protoplanetary disc formation. Recently, there have been detections of spiral features in young, embedded objects that appear consistent with GI-driven structure. It is crucial to perform hydrodynamic and radiative transfer simulations of gravitationally unstable discs in order to assess the validity of GIs in such objects, and constrain optimal targets for future observations. We utilise the radiative transfer code \lime\ to produce continuum emission maps of a $0.17\,\mathrm{M}_{\odot}$ self-gravitating protosolar-like disc. We note the limitations of using \lime\ as is and explore methods to improve upon the default gridding. We use \casa\ to produce synthetic observations of 270 continuum emission maps generated across different frequencies, inclinations and dust opacities. We find that the spiral structure of our protosolar-like disc model is distinguishable across the majority of our parameter space after 1 hour of observation, and is especially prominent at 230\,GHz due to the favourable combination of angular resolution and sensitivity. Disc mass derived from the observations is sensitive to the assumed dust opacities and temperatures, and therefore can be underestimated by a factor of at least 30 at 850\,GHz and 2.5 at 90\,GHz. As a result, this effect could retrospectively validate GIs in discs previously thought not massive enough to be gravitationally unstable, which could have a significant impact on the understanding of the formation and evolution of protoplanetary discs.

\end{abstract}


\section{Introduction}

Simulations have shown the surrounding protoplanetary disc can contain a mass comparable to that of the central protostar in newly formed systems \citep[e.g.][]{Machida&Matsumoto2011}. In this scenario, gravitational instabilities (GIs) can form in the disc and drive global spiral waves. As these spiral waves grow, they can produce shocks that heat the disc material locally \citep[e.g.][]{Harker&Desch2002, Boley&Durisen2008, Bae&Hartmann2014}, which has a significant effect on the chemical evolution of some species as \citet{Ilee&Caselli2011} and \citet{Evans&Ilee2015}, hereafter \citetalias{Evans&Ilee2015}, have shown. As the dust emission within discs is dependent on the temperature, the spiral shocks should produce a flux contrast between the arm and inter-arm regions. Therefore, continuum observations of GI-driven spiral structure in young, embedded systems would be of great importance as the formation and evolution mechanisms of protoplanetary discs are still uncertain.

\smallskip

There has been a continuous advancement in observational instruments which has resulted in an unprecedented amount of protoplanetary disc images at millimetre wavelengths. However, because spatially resolving the youngest, most embedded sources (late Class 0/early Class I) at au scales (the expected size of GI-driven spirals) remains very challenging, most observations to date have focused on more evolved discs (Class II). Spiral features have been detected in Class II discs \citep[e.g.][]{Muto&Grady2012, Benisty&Juhasz2015}, but \citet[][]{Dong&Hall2015} raise several important points that challenge the validity of GI-driven spiral arms in more evolved objects. Firstly, derived disc masses are not typically high enough to induce gravitational instabilities. Secondly, the accretion rates in observed Class II objects do not appear consistent with the accretion rates in their simulations of gravitationally unstable discs. Thirdly, the observed features are probably beyond the critical radius where discs are believed to fragment. \citet[][]{Dong&Hall2015} conclude by offering planet-disc interactions as a more credible origin for the spiral features, but it should be noted that there are counterarguments to each of the aforementioned points, and even planet-disc interactions may not explain observations to satisfaction \citep{Richert&Lyra2015}.

\smallskip

Since the Atacama Large Millimetre / submillimetre Array (\alma) became operational, the sensitivity of millimetre observations has increased, allowing us to peer deep into younger systems for the first time. As a result, \citet{Perez&Carpenter2016} have detected spirals originating from the bulk of the disc surrounding Elias 2-27 that appear consistent with gravitational instabilities \citep{Tomida&Machida2017, Meru&Juhasz2017}. Moreover, \citet{Tobin&Kratter2016}, have detected spiral features in a Class 0 object that also appear likely to have originated from gravitational instabilities, albeit in a fairly complex multiple star system where the spirals may be produced by gravitational interaction between the nearby young stellar objects. 

\smallskip

In light of these recent discoveries, it is currently a very exciting time in protoplanetary disc observations. However, even though spirals are beginning to be detected in real discs, it remains important to simulate observations, where all the properties are known {\textit{a priori}}, in order to assess the validity of GIs in young discs. Furthermore, simulated observations will allow the constraint of the observational parameters necessary for modern technologies such as \alma\ to unequivocally detect GI-driven spiral features. Several authors have reported that GI-driven spirals should be detectable with \alma\ in discs with a range of masses and sizes \citep[e.g.][]{Cossins&Lodato2010, Douglas&Caselli2013, Dipierro&Lodato2014}, and we build upon this repository of results by synthesising observations of the hydrodynamic simulation used in \citetalias{Evans&Ilee2015}, which is a gravitationally unstable $0.17\,\mathrm{M}_{\odot}$ disc surrounding a $0.8\,\mathrm{M}_{\odot}$ protostar. This system may be analogous to our early Solar System, so the synthetic detection of spiral structure in this case would be of particular significance.

\smallskip

An important result derived from observations of protostar systems is the mass of the surrounding protoplanetary disc as it is a pivotal quantity in understanding disc formation and evolution. Estimates from observations of the nearby Taurus star-forming region have determined that Class II disc masses range between 0.0003--0.06\,$\mathrm{M}_{\odot}$ \citep{Andrews&Rosenfeld2013} and masses of Class 0 discs range between 0.001--0.6\,$\mathrm{M_J}$ \citep[see][Table 1]{Greaves&Rice2011}. These estimates are based on the assumption of vertically isothermal disc structure and optically thin dust emission, which is perhaps appropriate for lower mass discs, although even this is debatable as masses derived from accretion rates tend to be higher. However, for younger and more embedded discs these assumptions are more likely to be inaccurate for a number of reasons. Firstly, shocks driven by GIs heat and lift material from the midplane, hence the vertical structure is far from isothermal \citep[see][]{Boley&Durisen2006, Evans&Ilee2015}. Secondly, due to the surrounding envelopes and high densities, Class 0/I discs are likely optically thick even at millimetre wavelengths \citep[e.g.][]{Miotello&Testi2014}. As a result, the observed flux from very young systems would only be tracing a fraction of the actual disc mass. This has implications for existing and future studies as derived masses could be underestimated, which could retrospectively validate GIs in discs previously thought not to be gravitationally unstable.

\smallskip

In this paper, we take the radiative hydrodynamic model of a gravitationally unstable disc from \citetalias{Evans&Ilee2015} and use it to investigate the observability of GI-driven spiral structure in dust continuum emission using ALMA. We establish the optimal parameters with which to perform the \lime\ radiative transfer, a crucial step for our next study on the observability of molecular species, for which we calculate the chemical evolution in Paper I (Evans et al., in prep.). We produce ALMA synthetic observations of our disc model at different frequencies and inclinations by implementing a range of dust opacities in the radiative transfer calculations. Our aim here is to provide an insight into how sensitive the interpretations of observations of gravitationally unstable discs can be to the assumed dust opacities, as the values for young, embedded objects are poorly constrained. We then compare observationally derived disc masses with the actual disc mass and compare assumed dust temperatures with the actual dust temperatures. We also provide a comparison between synthetic observations derived from \lime\ images to those derived from another existing radiative transfer code, {\sc radmc-3d} \citep{Dullemond&Juhasz2012}, because it is important to ensure there is consistency between different radiative transfer codes in the literature.

\smallskip

In Section \ref{sec:LIME} we briefly explain how \lime\ can be used to produce intensity maps and explore methods to improve upon the base \lime\ setup, which we refer to as `vanilla' \lime. These include: changing the weighting parameters; changing the number of grid points; restricting grid point positioning based on the optical depth; and the effect of averaging multiple runs. In Section \ref{sec:obs} we produce synthetic observations of our disc model across a large parameter space and use the flux density to determine the disc mass at each combination of parameters. We then compare the observationally determined disc masses to our actual simulation mass to assess the validity of this commonly used observational method. Finally, in Section \ref{sec:conclusions} we present our conclusions and discuss future research.

\section{Producing intensity maps with \lime}
\label{sec:LIME}

\subsection{Disc model}

The disc model we use to produce continuum emission maps is a snapshot of the lower mass disc featured in Paper I, taken at the end of the simulation, $t$ $\approx$ 2050\,yr, once the spiral features are fully developed and the disc is in a rough balance between heating and cooling processes, i.e. the disc is self-regulated and not fragmenting. Details of the hydrodynamic simulation can be found in \citetalias{Evans&Ilee2015}, which we briefly describe here. The disc has a mass of $0.17\,\mathrm{M}_{\odot}$, spans approximately 50\,au in radius, and the simulation was performed using 3D radiative hydrodynamics. The heating and cooling mechanisms consist of radiative energy losses, $PdV$ work, viscous dissipation and irradiation by a central protostar. The regular grid consists of consists of 512, 512, and 64 cells with 0.25\,au resolution in the $x$, $y$ and $z$ directions, respectively.

\smallskip

The dust model adopted in the hydrodynamic simulation assumes \citet{D'Alessio&Calvet2001} opacities with a grain size distribution $n(a)$ = $n{_0}a^{-3.5}$ and $\alpha_{max}$ = 1\,mm to account for grain growth \citep{D'Alessio&Calvet2006}. We use a gas-to-dust mass ratio of 100 and the dust grains are assumed to be thermally coupled and well mixed with the gas since we do not expect significant dust settling in such turbulent systems, and the computed Stokes number for mm grains is only greater than unity at the very edge of the disc.

\smallskip

We do not include an envelope in our radiative hydrodynamic simulation as we are only focusing on the detectability of GI-driven spiral structure. In reality, a surrounding envelope may shield the disc from external radiation that can otherwise diminish the spiral structure \citep[e.g.][]{Cai&Durisen2008, Kratter&Murray-Clay2011}, but here we are assuming our disc is already well-shielded, i.e. heating from the interstellar radiation field is neglected. Furthermore, infall from the envelope onto the disc can significantly affect the spiral structure \citep[e.g.][]{Harsono2011}, but as the consequence appears to be an enhancement in the contrast between arm and inter-arm regions, neglecting this interaction will only be significant for our conclusions if we do not detect GI-driven spirals.

\smallskip

We do, however, incorporate an envelope into the radiative transfer step of our modelling when producing continuum emission maps. In order to appropriately describe the environment that a young protoplanetary disc is embedded within, we take this envelope to be a $10\,\mathrm{M}_{\odot}$ contracting Bonnor-Ebert sphere \citep[see][]{Keto&Caselli2014}. Moreover, whilst radiative transfer codes often self-consistently calculate the dust temperature, this only considers radiative heating. Instead, we utilise the dust temperatures from the radiative hydrodynamics code, which combines flux-limited diffusion with ray-tracing in the vertical direction \citep[see][]{Boley&Durisen2007}, in order to account for the viscous and shock heating in our model. As a result we only need to perform raytracing on our disc model to obtain continuum emission maps.

\begin{figure}
    \includegraphics[width = 0.475\textwidth]{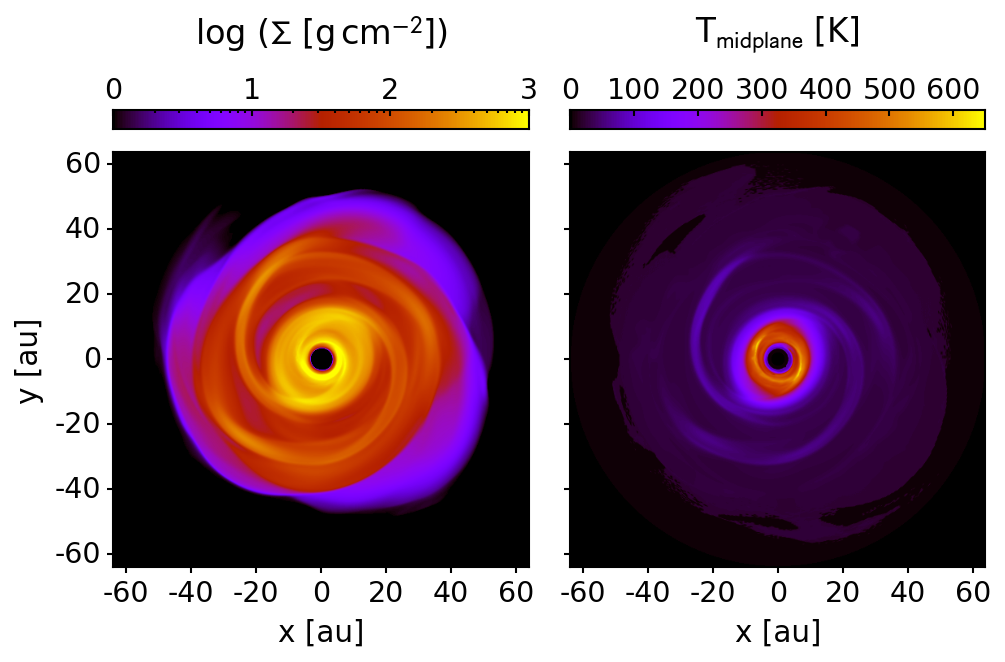}
    \caption{Surface density map of the disc model (left) and midplane temperature map (right), showcasing the non-axisymmetric spiral structure in the disc model.}
    \label{fig:ncd-temp}
\end{figure}

The surface density map and midplane temperature of the disc snapshot we utilise are shown in Figure \ref{fig:ncd-temp} to enable visual comparison to the continuum emission images presented in this paper. 

\subsection{`Vanilla' \lime}

We use the Line modelling Engine \citep[\textsc{lime};][]{Brinch&Hogerheijde2010}, which calculates continuum emission and line intensities from a weighted Monte Carlo sampling of an input 3D model. Here we provide a brief description of how \lime\ operates, but we refer the reader to \citet{Brinch&Hogerheijde2010} and \citet[][Section 2.3]{Douglas&Caselli2013} for more detailed descriptions.

\smallskip

The input model contains density and temperature information, and also abundance and velocity information if considering line emission. \lime\ then selects an $x, y, z$ position randomly, and if a selection criterion is met, which is dependent on the density by default, this point is added to the grid. Once the grid is constructed, \lime\ calculates the appropriate parameter values at each point; the method used to achieve this, such as nearest point interpolation, linear interpolation, or something more advanced, is defined by the user in the model file. The constructed grid is then smoothed via Lloyd's algorithm (Lloyd 1982, Springel 2010) in order to ensure the distance between points is comparable to the local separation expectation value whilst still maintaining the underlying model structure. The grid is then Delaunay triangulated and if non-LTE line emission is being considered, photons are propagated along the Delaunay lines and the radiative transfer equation is solved at each grid point via an iterative process. Finally, once convergence is reached, or if only continuum emission is being considered, \lime\ ray-traces lines of sight through the model in order to produce intensity maps.

\smallskip

\lime\ is in continual development, has been hosted on GitHub since v1.3\footnote{https://github.com/lime-rt/lime}, and the available documentation is rapidly improving. For now, \lime\ offers the user the ability to specify a whole array of parameters that affect the resultant intensity maps, which can be complex for first time users. In order to highlight potential pitfalls, we ran \lime\ v1.6 as is, which we refer to as `vanilla' \lime, for our disc model using $2.5 \times 10^4$ grid points; we use this number of grid points initially as the documentation recommends `between a few thousands up to about one hundred thousand' and it is also the number of grid points used in \citet{Douglas&Caselli2013}. 

\begin{figure*}
    \includegraphics[width = 0.95\textwidth]{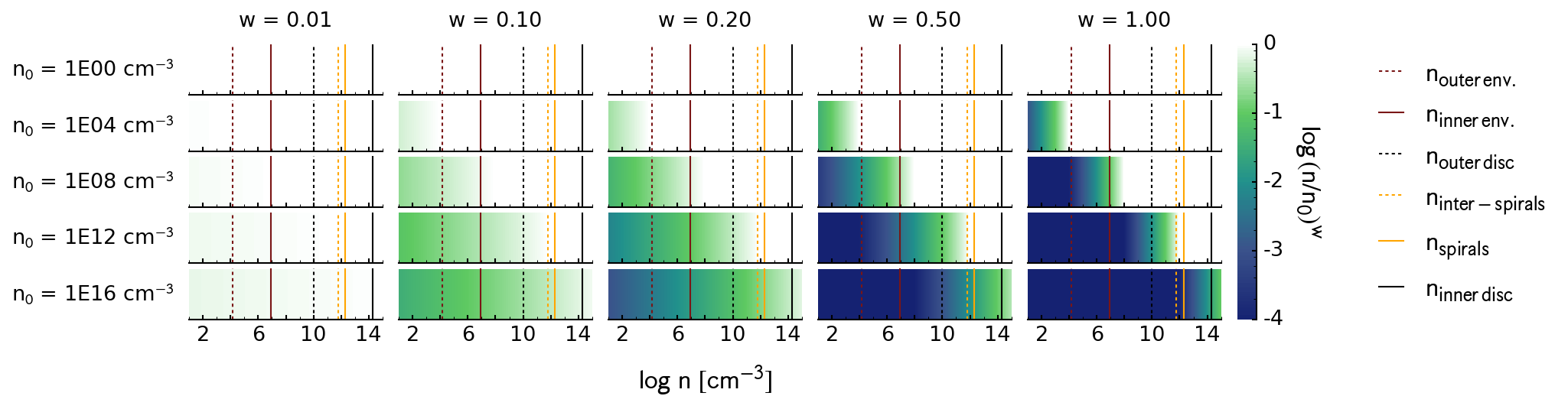}
    \caption{Probability of grid point selection as the density weighting parameters, $n_0$ and $w$, are varied. The vertical lines correspond to approximate number densities within various features of the disc model, as labelled.}
    \label{fig:weightingbars}
\end{figure*}

\begin{figure*}
    \includegraphics[width = 0.95\textwidth]{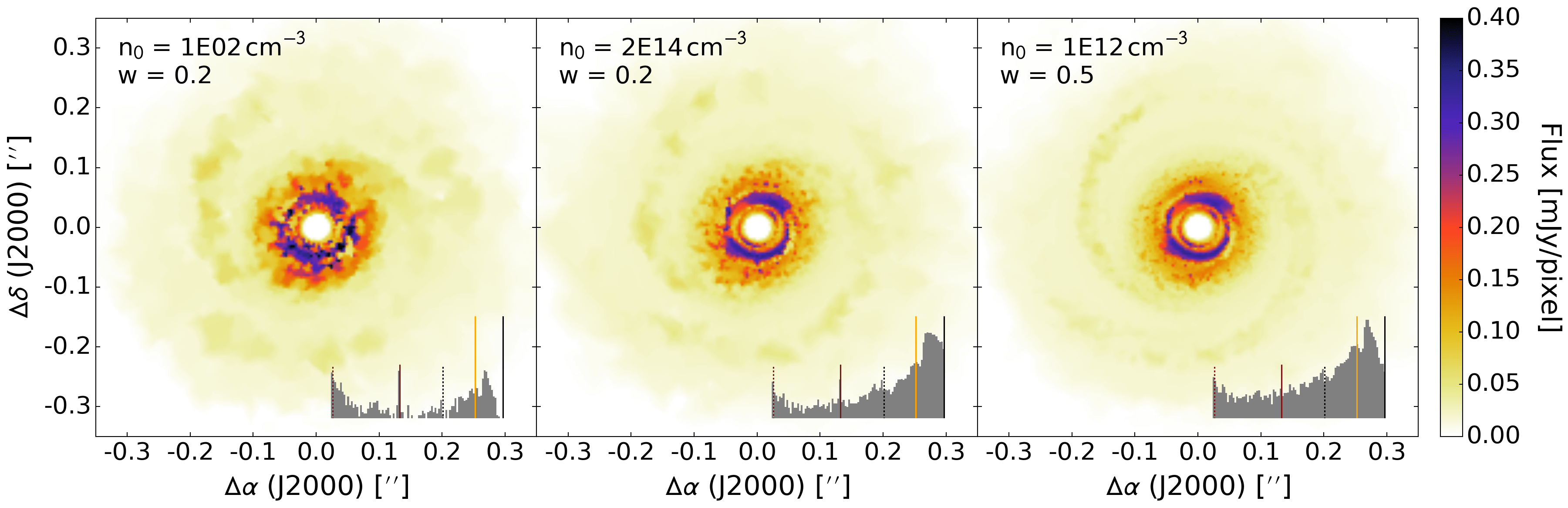}
    \caption{Comparison between 300\,GHz flux images produced by \lime\ using varied $n_0$ and $w$ parameters, demonstrating that \lime\ images are sensitive to the weighting imposed. The inset figures show the distribution of number densities of the grid points, confined within the disc region, that produce the corresponding image, with the vertical lines corresponding to the key in Figure \ref{fig:weightingbars}.}
    \label{fig:weightingcompare}
\end{figure*}

The leftmost panel of Figure \ref{fig:weightingcompare} shows the 300\,GHz continuum emission map for our disc model \citepalias[see][]{Evans&Ilee2015} when using vanilla \lime, and comparing to Figure \ref{fig:ncd-temp} it is immediately apparent that the non-axisymmetric structure is very poorly resolved. This is because the default weighting routine sets a normalisation density at the model inner boundary, which in our model is within the inner hole. This means we are essentially omitting any density weighting when using vanilla \lime. Note, we adopt a frequency of 300\,GHz for the rest of Section \ref{sec:LIME}.

\subsection{Weighting of grid points}
\label{sec:weighting}

In order to procure a more accurate flux image we must amend the density weighting function to be appropriate for our model. \lime\ randomly distributes points during the grid building routine and only selects the point if it passes a criterion
\begin{equation}
    a < \left({\frac{n}{n_0}}\right)^w
\end{equation}
where $a$ is a random number generated between 0 and 1 each time a point is selected, $n$ is the density of the grid point, $n_0$ is a normalisation density and $w$ is an exponent. The normalisation density was fixed to the inner boundary density in \lime\ v1.5 and earlier, but since v1.6 the user should now set the parameter ${n_{0}}^{w}$ in the model file. If the user does not set this parameter however, then the old default is restored. This can be a poor selection criterion if, for example, the model has an inner hole or if the user would like to focus on features with a lower density than the central density. We demonstrate this effect in Figure \ref{fig:weightingbars}, which shows the probability of point selection, $\mathrm{log(n/n_0)^w}$, as $n_0$ and $w$ are varied, with features of our disc model overlaid. As can be seen, if the reference density, $n_0$, is increased then the likelihood of point selection at low densities becomes increasingly small, and if the exponent, $w$, is increased then the range where points are likely to be selected contracts. For our purposes we want a high probability of point selection at typical spiral arm densities, indicated by the solid orange vertical lines, but also with a significant percentage of points extending to the outer disc regions, with an approximate density denoted by the dotted dark gray lines. Therefore, from Figure \ref{fig:weightingbars} we determine that \nzero\ and $w = 0.5$ are optimal choices for our disc model. Note that this choice of parameters results in a small fraction of points being positioned within the envelope. This is not an issue, however, because the maximum envelope density is lower than the bulk of the disc model. As a result, the mean free path within our simplistic envelope model is ubiquitously larger than the envelope size and hence we only need a small fraction of points to treat the emission accurately. In reality, the innermost part of the envelope around embedded protoplanetary discs may be very dense, which could affect our conclusions in the outer disc regions.

\smallskip

We ran vanilla \lime\ for our disc model, which equates to using \nnone\ and $w = 0.2$. We then ran \lime\ using \nfull\ and $w = 0.2$ to simulate the intention of the default reference density value assignment. Finally, we ran \lime\ using \nzero\ and $w = 0.5$ that, as aforementioned, we determined from Figure \ref{fig:weightingbars}. Figure \ref{fig:weightingcompare} shows the flux image outputs of these runs and as can be seen, if $n_0$ is set much lower than the typical disc density then essentially no weighting is applied, apart from excluding the inner hole. In this case there is poor sampling of the densest features and a large percentage of points are positioned in regions of low density. If, on the other hand, we set $n_0$ to the highest density in our model, as shown in the middle panel, then density weighting is applied and the disc spirals and inner region are sampled much more thoroughly. However, if we compare this result to the result in the right panel, where $n_0$ is set at the approximate spiral density, we can see that using too high a reference density undersamples the spiral features comparatively. This is because, as Figure \ref{fig:weightingbars} shows, the probability of point selection at the density of the spiral arms is reduced as $n_0$ increases. Indeed the panel on the right, with \nzero\ and $w = 0.5$, showcases a type of `Goldilocks' regime for our disc model, in which we sample the disc spirals in an optimal manner. Note, though, that the trade-off is a smaller percentage of points at the highest densities, which in our model corresponds to the innermost disc, as indicated by the inset histogram in the right panel of Figure \ref{fig:weightingcompare}. Therefore, the optimal choice of $n_0$ is dependent on the features of interest in the model. For our purposes we use \nzero\ and $w = 0.5$ and refer to this setup as our standard sampling hereafter.

\subsection{Number of grid points}
\label{sec:numpoints}

Due to the Monte Carlo nature of the \lime\ gridding, two runs using identical inputs will likely produce differing images as the grid points will be positioned differently. This effect obviously decreases as the number of grid points is increased, so in order to circumvent this issue, the user can simply use a large number of grid points. For continuum images this is not a particularly significant issue because the computational cost is relatively low. However, for line images, because each molecular energy level population must be computed for each grid point within each iteration, the computational cost increases dramatically with the number of grid points. Therefore, even though this paper focuses on continuum images, deducing the minimum number of grid points we can use and still obtain accurate images in the continuum is a worthwhile endeavor as the results could be applicable to line images.

\begin{figure}
    \includegraphics[width = 0.475\textwidth]{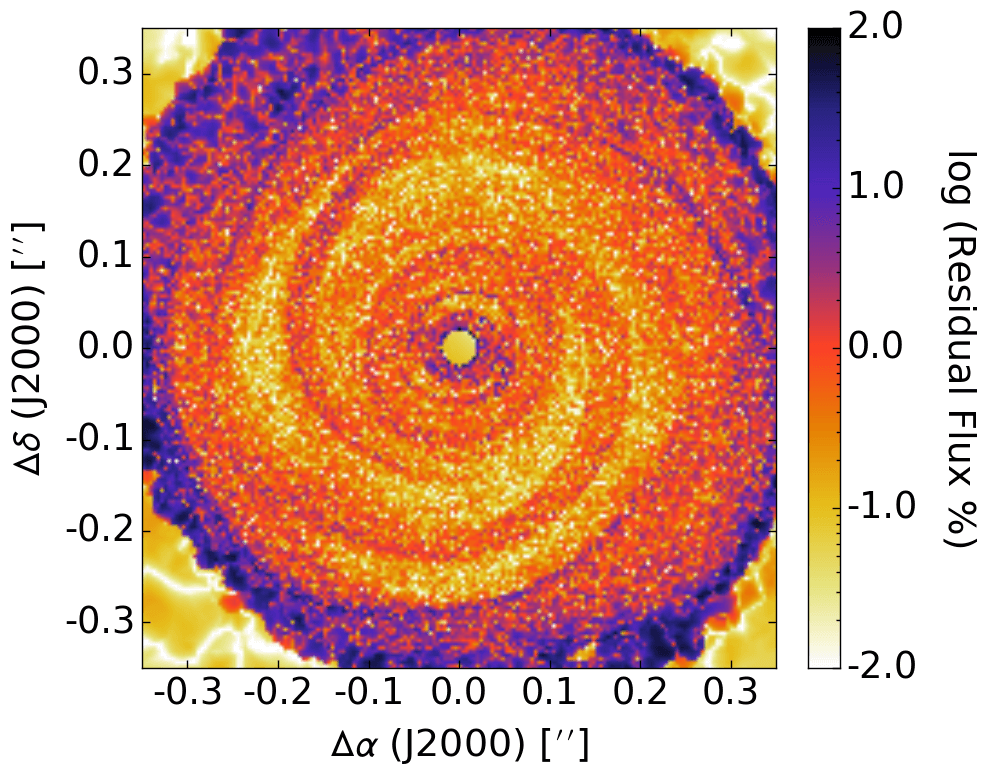}
    \caption{Residual flux between two \lime\ runs using identical inputs with $2 \times 10^6$ grid points and standard sampling. We define this residual as $|$(f$_1$-f$_2$)$|$/(f$_1$+f$_2$), where $f$ denotes flux per pixel.}
    \label{fig:residual2000000}
\end{figure}

To achieve this we adopt the weighting parameters discussed in Section \ref{sec:weighting} and produce continuum images with an increasing number of total grid points until the residual between two runs using identical inputs met a sufficient level. We set this level to 5 per cent because this is smaller than the errors expected in observations. We find that the majority of the residual flux is smaller than our threshold level when using $2 \times 10^6$ points, as Figure \ref{fig:residual2000000} shows. Note that the large residual differences seen towards the edges of Figure \ref{fig:residual2000000} are where the model transitions from the disc to the envelope and the abundance of points drops off sharply. As we are focusing on the spiral structures we omit this region in future analyses.

\begin{figure*}
    \includegraphics[width = 0.95\textwidth]{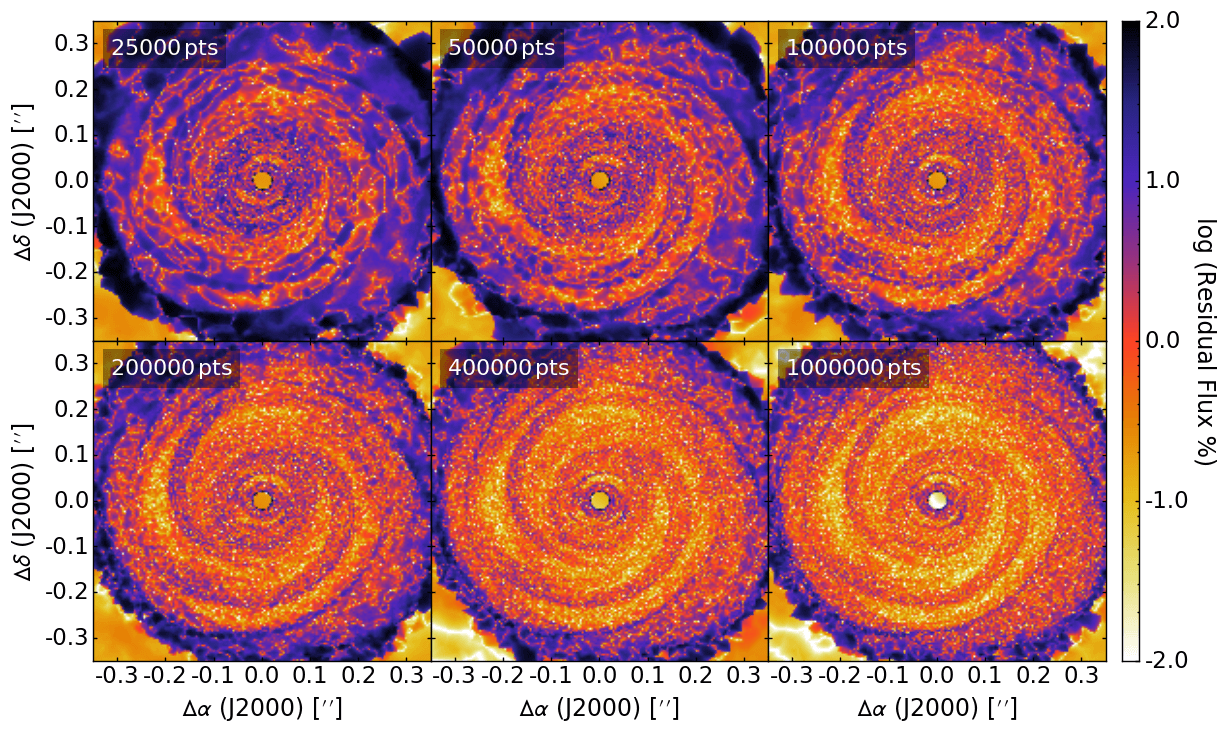}
    \caption{Residual flux between standard \lime\ runs using the indicated number of grid points and $2 \times 10^6$ grid points, with standard sampling. We define this residual as $|$(f$_1$-f$_2$)$|$/(f$_1$+f$_2$), where $f$ denotes flux per pixel.}
    \label{fig:pointsresidualcompare}
\end{figure*}

By comparing the images produced using fewer number of grid points with this `canonical' result, Figure \ref{fig:pointsresidualcompare} shows that increasing the number of grid points increases the accuracy of the residual images. This is obviously a trivial result, but more importantly, Figure \ref{fig:pointsresidualcompare} indicates that a sufficiently accurate image, which we have defined prior as the majority of the disc having a residual flux lower than 5 per cent, can be produced using $2 \times 10^5$ grid points. This is a factor of ten lower than our `canonical' result, and hence affords us a significant reduction in computational cost when producing continuum images. Not only is this a benefit in reducing computational time, but it also reduces memory consumption which allows more instances of \lime\ to be run simultaneously, which is utilised in Section \ref{sec:averages}.

\smallskip

Although the results demonstrated in Figure \ref{fig:pointsresidualcompare} are expected, it is important to understand precisely why an increase in grid points affords an improvement in the image accuracy. The emission of photons is governed by the mean free path, which is a statistical average of the distance a photon will travel before interacting. If the grid point separation is larger than the local mean free path, then important photon interactions will be omitted. Hence the resultant flux images will be inaccurate.

\smallskip

The mean free path is given by $l = 1/\alpha_{\nu}$, where $\alpha_{\nu}$ is the absorption coefficient at a particular frequency. For dust-dominated continuum emission $\alpha_{\nu}$ = $\kappa_{\nu}\rho_d$, where $\kappa_{\nu}$ is the dust opacity at a specific frequency and $\rho_d$ is the dust mass density. We know the $\kappa_{\nu}$ value as it is contained within a dust opacity file read by \lime, and we also have the gas mass density information within our hydrodynamic simulation, which we convert to dust mass density using the commonly adopted value of 100 for the gas-dust mass ratio. Using these quantities we calculate mean free path values throughout our disc model in an analytical fashion. Next, we straightforwardly obtain the grid point separation information as \lime\ v1.6 has the ability to output the grid point positions and nearest neighbour distances. Finally, we interpolate the nearest neighbour distances to a coordinate grid matching our mean free path data cube and compare these length scales.

\begin{figure*}
    \includegraphics[width = 0.95\textwidth]{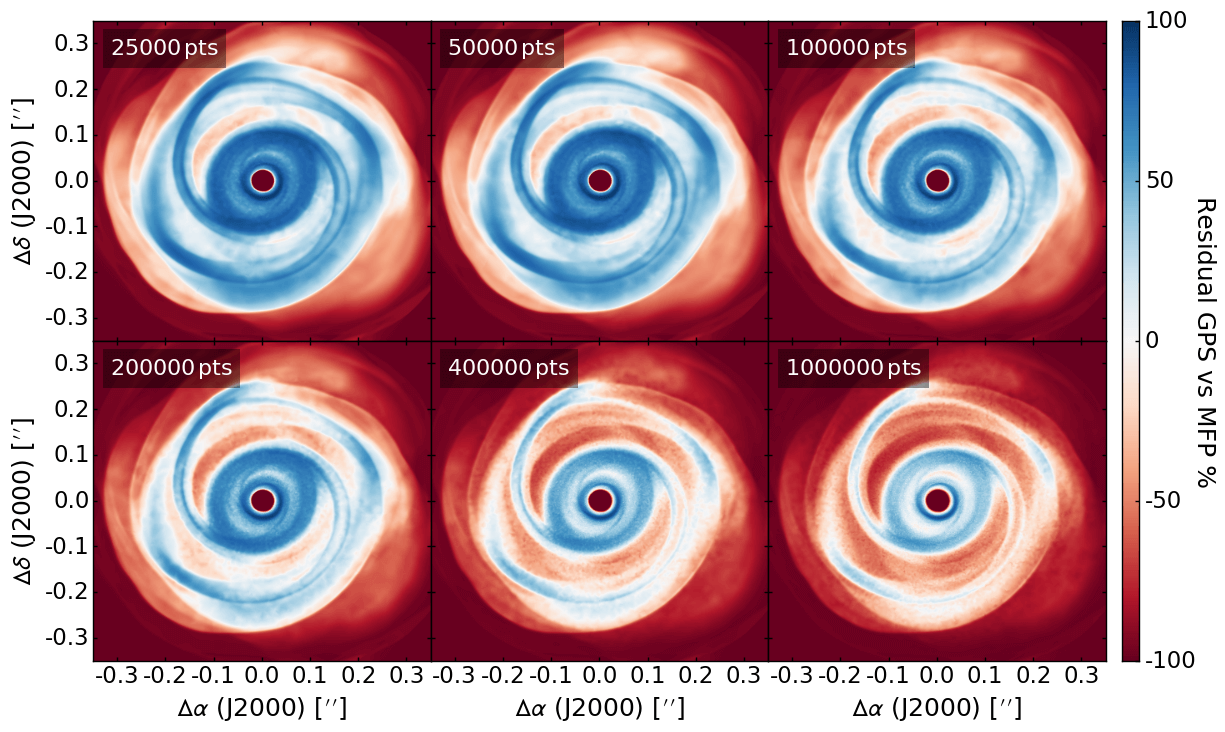}
    \caption{Residual of grid point separation (GPS), $s$, and mean free path (MFP), $l$, at the disc midplane for \lime\ runs using the indicated number of grid points and standard sampling. We define this residual as (s-l)/(s+l).}
    \label{fig:gpsvsmfp}
\end{figure*}

We plot the comparison between the mean free path and grid point separation across the $x-y$ plane at $z = 0$ for each of the runs shown in Figure \ref{fig:pointsresidualcompare}. The results are shown in Figure \ref{fig:gpsvsmfp} and demonstrate that by increasing the number of grid points, the disparity between these two length scales across the majority of the disc is reduced. However, even when using $1 \times 10^6$ points there is still an appreciable discrepancy between the mean free path and grid point separation within the spiral arms and inner disc regions. This is not surprising because these are the densest regions of our disc model and hence have the smallest local mean free path values. Fortunately, this is not a particularly large issue because the contribution to intensity in these densest, most optically thick regions is small. In fact, we can account for this phenomenon and optimise our sampling routine further.

\subsection{Optical depth surface}
\label{sec:tausurface}

In optically thick regions, such as those expected in embedded protoplanetary discs, the contribution to observed intensity is low because $I_{\nu,0} \propto e^{-{\tau}}$. Therefore, we designate a threshold of $0.05I_{\nu,0}$, below which point we assume the emission becomes negligible as this is less than typical observational errors concerning embedded objects. This means we should only need to place grid points up to a surface where the optical depth reaches $\tau$ = -ln(0.05) = 3 in order to produce an accurate image. We refer to this surface as the optical depth surface hereafter. For reference, the photosphere of a star is defined at $\tau$ = 2/3, hence we are likely still oversampling the densest regions of our disc model by adopting $\tau$ = 3 for our optical depth surface. However, determining the minimum optical depth one can use in grid point restriction and still obtain accurate images is dependent on dust opacity and frequency, which requires a dedicated study and as such is beyond the scope of this work.

\smallskip

\lime\ is capable of outputting several different image types, such as brightness temperature, flux and optical depth along the line-of-sight, which the user can specify; we adapt the way \lime\ outputs images so that the user can now choose to have multiple image types output for one set of image parameters, which has been incorporated into \lime\ v1.7.2. In order to implement our optical depth surface method\footnote{github.com/lolmevans/lime/tree/optical-depth-surface}, we add a new image output to \lime\ that allows us to determine where in our disc model this optical depth surface lies. This is achieved via an addition to the ray-tracing routine that records the $z$ position for each propagated ray once the optical depth has surpassed our specified value. The optical depth is calculated via $\tau$ = $\kappa_{\nu}\rho_d$ds, where $\kappa$ is the dust opacity at the specified frequency (300\,GHz), $\rho_d$ is the dust density and $ds$ is the distance between Voronoi cells after each iteration of the ray propagation. As the optical depth can jump significantly beyond our threshold value from one Voronoi cell to the next, we implement linear interpolation. We pass this value into an additional ray \textit{struct} parameter and then average over the number of rays that pass through each pixel in the same manner as the existing flux and optical depth images. The resultant image displays the height above or below the midplane of our desired optical depth surface across the entire disc model.

\begin{figure}
    \centering
    \includegraphics[width = 0.425\textwidth]{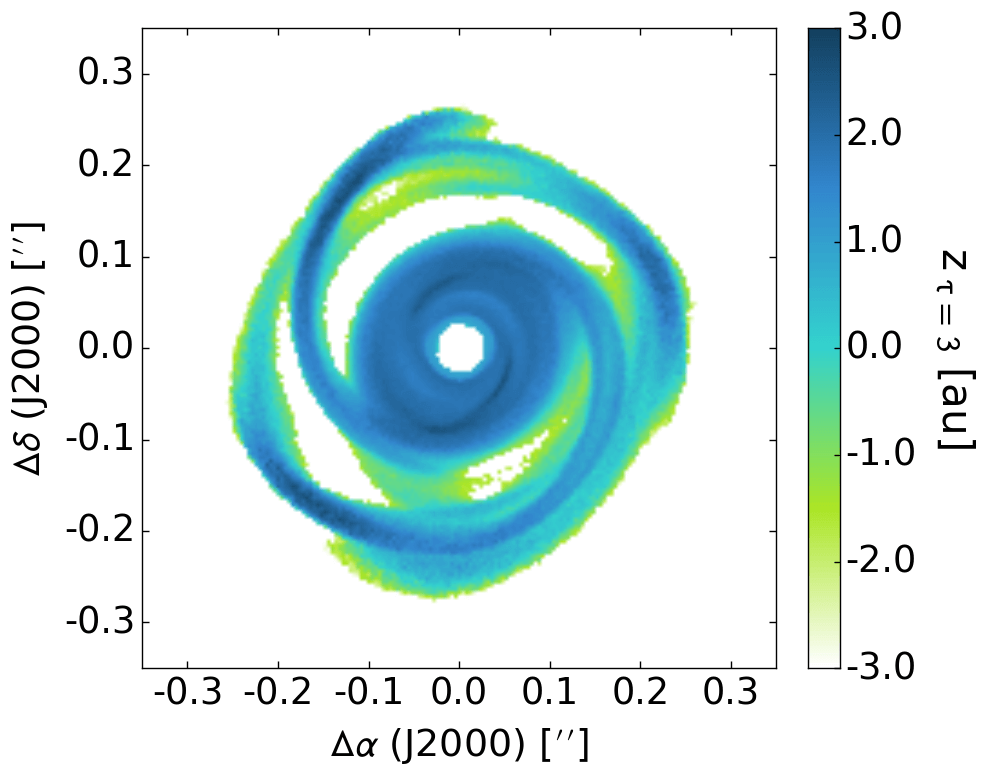}
    \caption{Surface across the disc model where $\tau = 3.0$ when viewed from above. Below this surface the contribution to intensity becomes insignificant.}
    \label{fig:tausurface}
\end{figure}

Figure \ref{fig:tausurface} shows the $\tau = 3$ surface for our disc model when viewed from above. The optical depth surface is significantly above the midplane within the spiral arms due to the large column densities within the spiral arms, and because the interaction between disc material and the spiral shocks is similar to a hydraulic jump (Boley \& Durisen 2006). Hence, material in the spiral arms is elongated vertically. Between the spiral arms, the optical depth surface lies below the midplane, which means we are peering through significantly more disc material in these regions.

\smallskip

As the contribution to the emission is less than $0.05I_{\nu,0}$ below the optical depth surface, using Figure \ref{fig:tausurface} we can see that omission of grid points below $z\,\approx\,3.0\,\mathrm{au}$ in the spirals, below $z\,\approx\,2.0\,\mathrm{au}$ across the inner disc and below $z\,\approx\,-1.0\,\mathrm{au}$ in the inter-arm regions is reasonable. Rather than use these approximate values, however, we amend the \lime\ gridding function to read in the disc optical depth surface and use this as a lookup table for the minimum $z$ position permissible during sampling. Therefore, if a point is selected with a $z$ position below the optical depth surface it is translated vertically to above the surface; the distance translated is proportional to the original distance from the midplane so that we effectively move the focus of points from around the midplane to above the optical depth surface only.

\begin{figure*}
    \includegraphics[width = 0.95\textwidth]{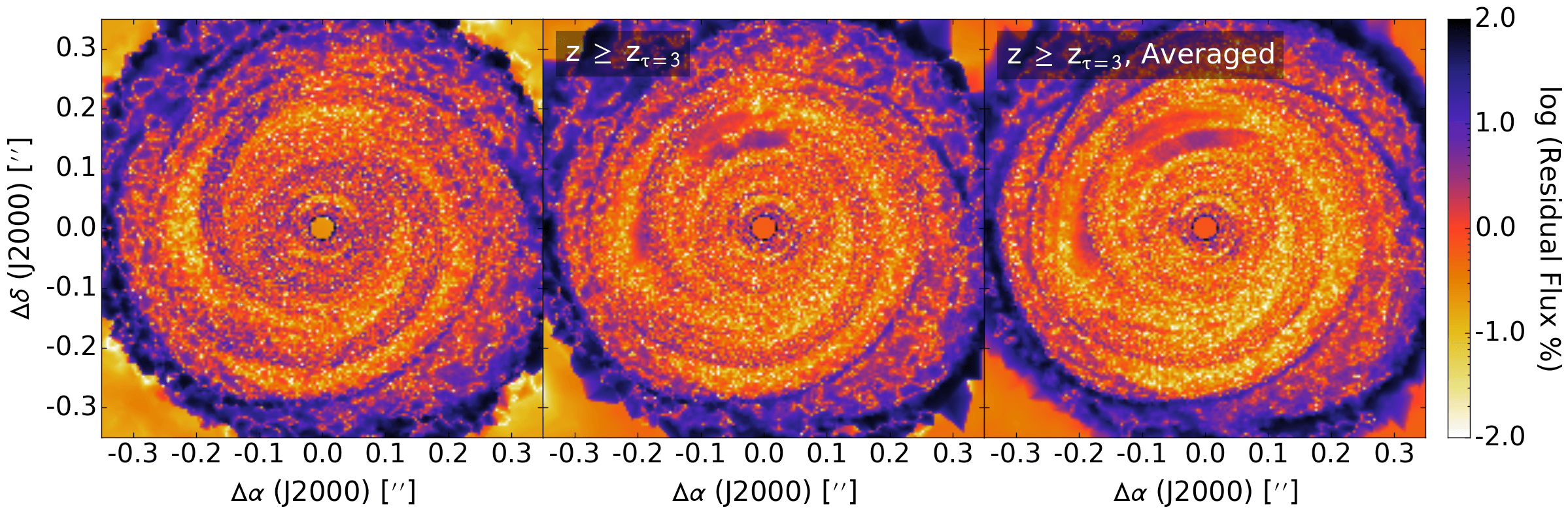}
    \caption{Residual flux between \lime\ runs using $2 \times 10^5$ grid points and different sampling methods, f$_1$, and $2 \times 10^6$ grid points with standard sampling, f$_2$. We define this residual as $|$(f$_1$-f$_2$)$|$/(f$_1$+f$_2$), where $f$ denotes flux per pixel. The panel on the left is the residual result for our standard sampling seen in Figure \ref {fig:pointsresidualcompare}. The middle panel shows the residual result when using an optical depth surface to constrain the positioning of grid points, which we refer to as our optimal sampling. The right panel shows the residual result for an average of eight optimally sampled \lime\ runs using identical input parameters.}
    \label{fig:zlimitedaverageresidualcompare}
\end{figure*}

We use this technique to reproduce the \lime\ image consisting of $2 \times 10^5$ grid points, calculate the residual from $2 \times 10^6$ points, and compare this to the bottom left panel of Figure \ref{fig:pointsresidualcompare}. The results are shown in the left and middle panels of Figure \ref{fig:zlimitedaverageresidualcompare} and as can be seen, limiting the grid points to above the optical depth surface affords an improvement in the image overall. In fact, when comparing to Figure \ref{fig:pointsresidualcompare} we can see that the effect is similar to doubling the number of grid points, but without the large increase in computational time; implementing our method has no significant effect on the runtime but doubling the number of grid points increases the runtime by 400 per cent. It should be noted that in order to implement our optical depth surface routine we need to run \lime\ initially to generate the optical depth lookup table. Despite this, however, our method is still more efficient than producing a continuum image with double the number of grid points. Therefore, this result is a proof of concept that positioning points more intelligently can result in more efficient radiative transfer calculations using \lime.

\smallskip

Although there is a general improvement visible between the left and middle panels of Figure \ref{fig:zlimitedaverageresidualcompare} there are also some regions where the residual has deteriorated. This is particularly evident in the prominent dark feature above the inner hole. However, the residual difference in this region is approximately 5 per cent, which is within the expected threshold of omitted emission when using our $\tau\,=\,3$ surface to constrain point positions. Furthermore, this residual difference equates to an absolute difference of 2\,K and is located within a weakly emitting region. As a result, this rather visually striking difference is entirely insignificant, but we offer a possible explanation for completeness: there is a particularly sharp density gradient in this region and hence our optical depth method is not treating this area properly. This then implies that the density weighting employed by \lime\ is not entirely accurate for our method and a better alternative could be a weight dependency on the density gradient. This, however, is beyond the scope of this paper and is left as a future task, especially considering the small residual difference present here.

\smallskip

The reason this method produces a more accurate image is entirely consistent with our results from Section \ref{sec:numpoints}, where we increase the total number of grid points, which reduces the grid point separation throughout the entire disc. In this section, however, we are increasing the number of grid points in the emitting regions using selective positioning. This therefore reduces the disparity between grid point separation and mean free path within the emitting regions only, i.e. at $z \geq z_{\tau=3}$.

\subsection{Averaging runs}
\label{sec:averages}

Multiple instances of \lime\ can be run simultaneously, limited only by the number of processor cores and memory available. The averaging of ten runs using identical inputs was performed in \citet{Douglas&Caselli2013} (see their Figure 6), which reduced image artifacts and smoothed their flux emission successfully. We adopt the same technique but use our optimal image setup (\nzero, $w = 0.5$, $2 \times 10^5$ points, $z \geq z_{\tau=3}$) and only average eight runs due to computational restraints. 

\smallskip

The comparison between a single run and the averaging of multiple runs using identical inputs is shown in the middle and right panels of Figure \ref{fig:zlimitedaverageresidualcompare}, and as can be seen, averaging affords us an improvement to the output image. Due to the Monte Carlo nature of the \lime\ gridding, averaging multiple identical runs effectively increases the point coverage. This is essentially identical in effect to increasing the number of grid points, which, as shown in Section \ref{sec:numpoints}, improves the image due to the disparity reduction between the grid point separation and mean free path. It is worth noting that the improvements seen between the middle and right panels of Figure \ref{fig:zlimitedaverageresidualcompare} are small in magnitude and likely much less than observational errors. However, because multiple instances of \lime\ can be run simultaneously, which affords a significant reduction in computation time compared to increasing the number of grid points, there seems to be no strong argument against using the average of the outputted images for further analysis. Furthermore, although we are only focusing on continuum images, which have an approximately linear increase in computation time with increased grid points, for line images this relationship is non-linear due to the iterative calculation of level populations. As a result, the benefit of averaging multiple runs using identical inputs should be emphasised when producing line emission images (Evans et al., in prep.).

\section{Synthetic observations}
\label{sec:obs}

\citet{Tobin&Kratter2016} detected spiral structure in a very young disc that hosts multiple protostars, which appears consistent with a gravitationally unstable disc that recently underwent fragmentation. Moreover, \citet{Perez&Carpenter2016} detected spiral structure in a single-protostar circumstellar disc that does appear consistent with GI-driven spiral arms \citep{Tomida&Machida2017}. Although the object featured in \citet{Tobin&Kratter2016} is not consistent with our model, and the simulated model of \citet{Tomida&Machida2017} does not match some features of the observation in \citet{Perez&Carpenter2016}, these results offer the first potential confirmation of GI-driven spiral structure in protoplanetary discs, and suggest more will follow in the near future. Assessing the validity of GIs in such objects is crucial and therefore we produce synthetic continuum emission maps of our GI-driven disc model in order to offer potential insight into this conundrum.

\begin{figure*}
    \includegraphics[width = 0.95\textwidth]{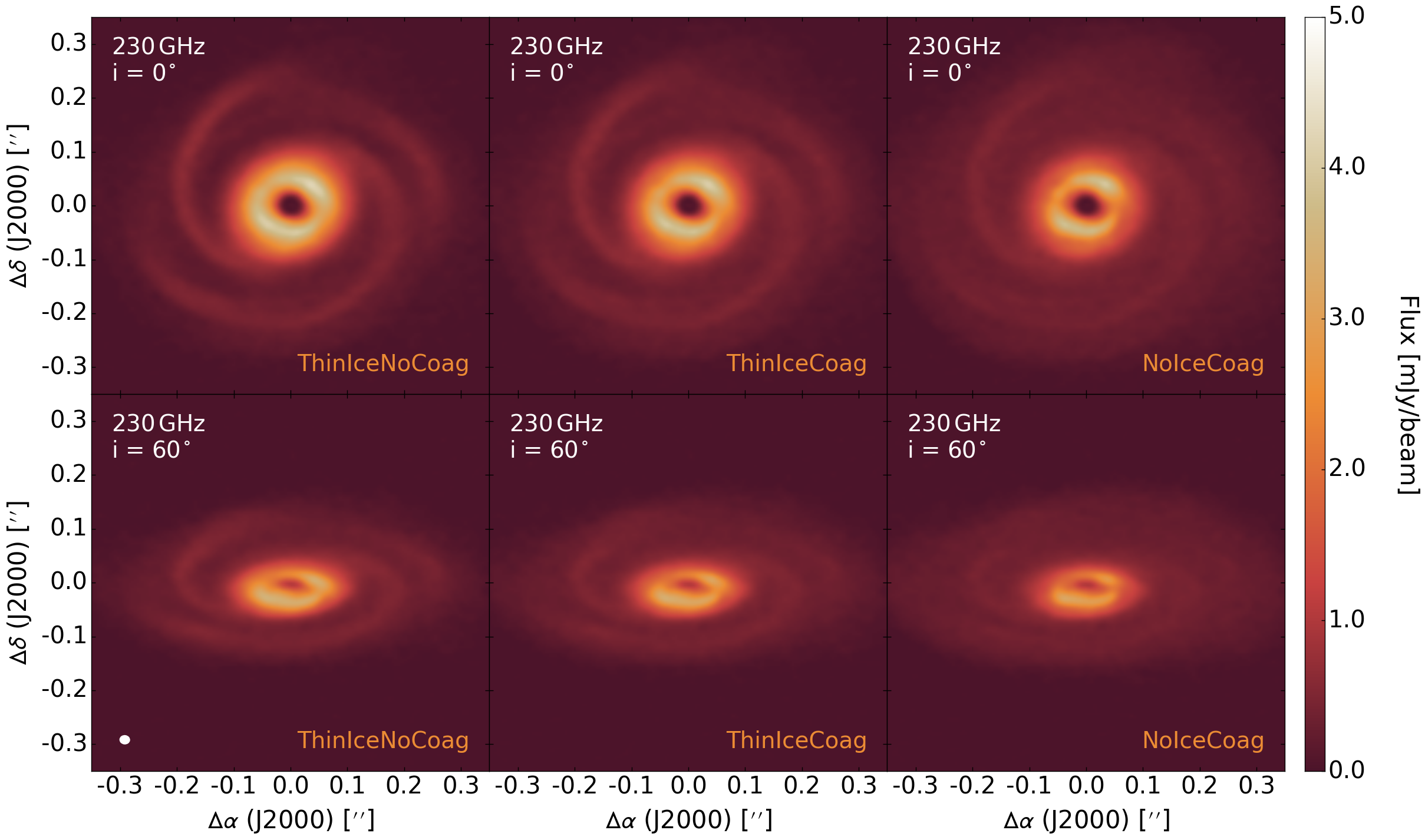}
    \caption{Continuum emission images of our disc model at 230\,GHz for three different dust opacity models and two different inclinations, synthesised using the \alma\ Cycle 5 antenna configuration. The white ellipse in the lower left indicates the size of the beam, which is 0.019$\,\times\,$0.017\,arcsec and is constant across all panels. The noise is approximately 15\,\textmu Jy per beam across the parameter space.}
    \label{fig:230GHzCycle5}
\end{figure*}

We produce 270 \lime\ images of our disc model across a 3-dimensional parameter space. We use $4 \times 10^5$ grid points and choose not to implement the optical depth surface method described in Section \ref{sec:tausurface} as we would need to produce a $\tau$ = 3 surface, and then run \lime\ again whilst implementing this surface. As we are only considering continuum emission using pre-computed dust temperatures, we only need to perform raytracing. Hence, the increase in image accuracy when adopting our optical depth surface model is offset by the significant increase in computational time from running \lime\ twice for each combination of dust opacity and frequency. Moreover, by comparing Figures \ref{fig:pointsresidualcompare} and \ref{fig:zlimitedaverageresidualcompare}, we can see that a single run of $4 \times 10^5$ points is comparable to our optimal image (produced using the average of eight runs implementing our optical depth surface method). Therefore, for this large parameter space study, we opt to use double the number of grid points without implementing our optical depth surface method in the interest of efficiency, whilst still producing accurate images for analysis. However, we emphasise that the purpose of Sections \ref{sec:tausurface} and \ref{sec:averages} is to demonstrate that the gridding routine \textit{can} be optimised for a particular model, which will be much more prevalent when producing line images as the computational time no longer scales linearly with the number of points due to the necessity in calculating level populations. We will investigate this in a forthcoming publication.

\smallskip

The parameter space we consider consists of varying frequencies, inclinations and dust opacities. We use five different frequencies commonly implemented in observations of protoplanetary discs, 90\,GHz, 230\,GHz, 300\,GHz, 430\,GHz and 850\,GHz, equivalent to 3.33\,mm, 1.30\,mm, 1.00\,mm, 0.70\,mm and 0.35\,mm, respectively, and the inclinations we use range from face-on (0$^{\circ}$) to edge-on (90$^{\circ}$) in 5$^{\circ}$ increments. We use three dust grain properties: coagulated grains with ice mantles (ThinIceCoag); coagulated grains without thin ice mantles (NoIceCoag); and non-coagulated grains with thin ice mantles (ThinIceNoCoag), all taken from \citet{Ossenkopf&Henning1994}, where `coagulated' refers to a coagulation after 10$^5$ years for a gas density of 10$^6$\,cm$^{-3}$. 

\subsection{Dust opacities}

\begin{table}
    \centering
    \begin{tabular}{cccccc}
        \cmidrule(l){2-6}
                                                      & \multicolumn{5}{c}{Frequency [GHz]}       \\ \cmidrule(l){2-6}
                                                      & 90       & 230    & 300    & 430    & 850     \\ \midrule
        \multicolumn{1}{C{2cm}}{Dust Grain Configuration} & \multicolumn{5}{c}{$\kappa_\nu$ $\rm [cm^2g^{-1}]$} \\ \midrule
        \multicolumn{1}{c}{ThinIceNoCoag}                & 0.112    & 0.509  & 0.782  & 1.488  & 5.827   \\ 
        \multicolumn{1}{c}{ThinIceCoag}                  & 0.200    & 0.896  & 1.369  & 2.593  & 9.938   \\
        \multicolumn{1}{c}{NoIceCoag}                    & 0.635    & 1.987  & 2.744  & 4.591  & 11.184  \\
        \multicolumn{1}{C{2cm}}{D'Alessio et al. (2001)} & 0.523    & 1.557  & 2.111  & 3.653  & 12.082  \\
        \bottomrule
    \end{tabular}
    \caption{Dust opacity values for different dust grain configurations across the frequency range used when deriving disc mass estimates.}
    \label{tab:freqkappas}
\end{table}

\begin{figure}
    \includegraphics[width = 0.475\textwidth]{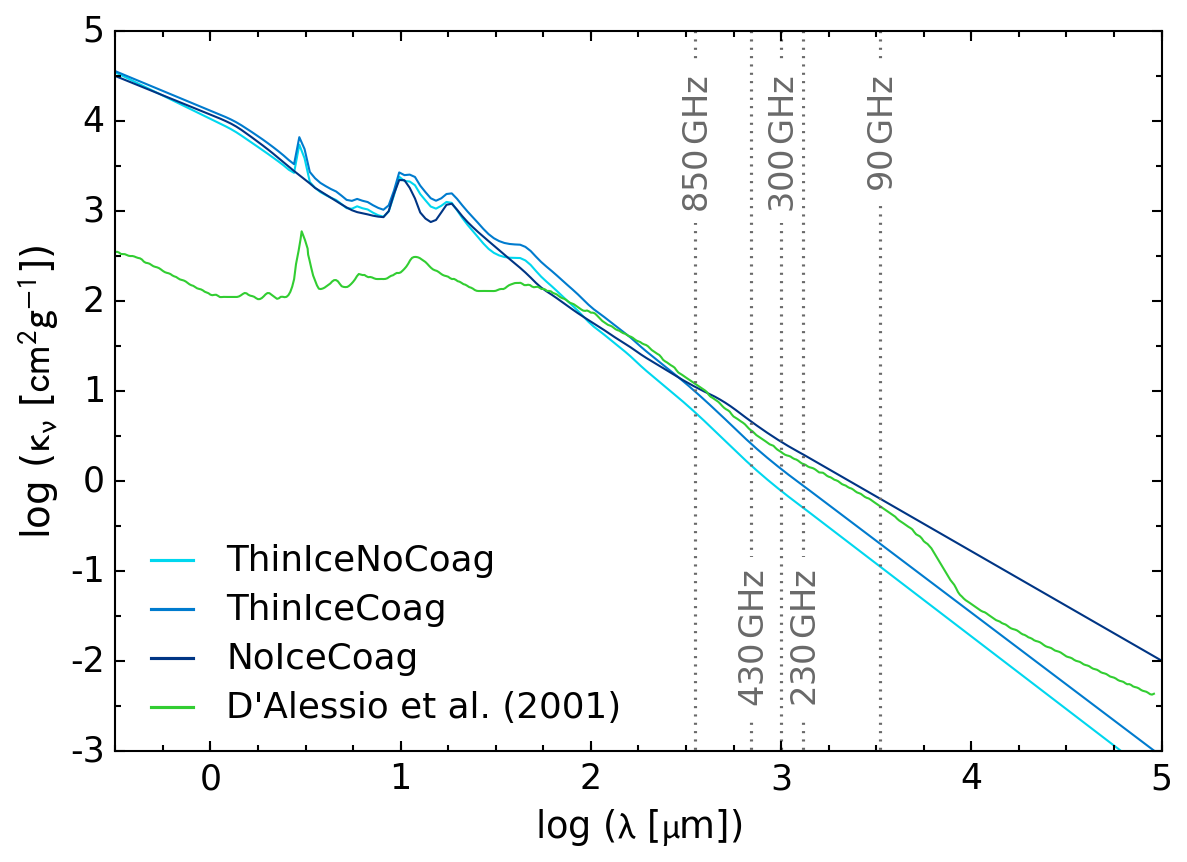}
    \caption{Dust opacity as a function of wavelength for the dust models used to produce synthetic observations of our disc model (`ThinIceNoCoag', `ThiniceCoag', `NoIceCoag'), and for the dust model used in the radiative hydrodynamic simulation \citep{D'Alessio&Calvet2001}.}
    \label{fig:opacities}
\end{figure}

The values of the dust opacities for the different dust models are shown in Table \ref{tab:freqkappas} at the frequencies we use, alongside the \cite{D'Alessio&Calvet2001} opacities used in the radiative hydrodynamic simulation. As can be seen, the `NoIceCoag' dust model is consistent (within a factor of 1.3) of the \citet{D'Alessio&Calvet2001} opacities across the observational frequency range we use (90--850\,GHz), whereas the `ThinIceCoag' and `ThinIceNoCoag' dust models diverge substantially at low frequencies. For this reason we acknowledge that synthetic observations produced using `ThinIceCoag' and `ThinIceNoCoag' dust opacities will not be entirely self-consistent with the disc simulation. However, we present these results as an exploration of the effects dust opacity can have on the observability of spiral structure. In truth, the evolution of a gravitationally unstable disc is affected by the heating and cooling processes \citep[see][]{Kratter&Lodato2016}, so repeating the radiative hydrodynamic simulation with, for example, the `NoIceCoag' dust model would most likely change the spiral structure in our disc model. In this case, as \autoref{fig:opacities} shows, the `NoIceCoag' opacities in \autoref{tab:freqkappas} are higher than the \citet{D'Alessio&Calvet2001} opacities at micron wavelengths, which would result in an increased cooling time. This suggests that the contrast between arm and inter-arm regions would be less pronounced \citep{Cossins&Lodato2009}, hence we would expect the detection of spiral structure to become more difficult. However, whilst self-consistently exploring the effects of dust opacity on the observability of spiral structure in a 3D radiative hydrodynamic simulation is an important topic of study, it is an extremely computationally expensive process and is beyond the scope of this paper.

\smallskip

Whilst the dust models in \autoref{tab:freqkappas} do not follow a simple power law $\kappa_\nu$ $\propto$ $\nu^\beta$ across our frequency range, we can extract $\beta$ values at specific frequencies to allow comparisons to values derived in the literature. For instance, $\beta_{230\,GHz}$ = 2.02, $\beta_{230\,GHz}$ = 1.64 and $\beta_{230\,GHz}$ = 1.10 for the `ThinIceNoCoag', `ThinIceCoag' and `NoIceCoag' dust models respectively. \citet{Miotello&Testi2014} find $\beta$ $\approx$ 0.5-1.0 in Class I objects which may suggest a higher level of grain growth than we consider. However, the relative contribution of envelope and disc to this value is not clear, and the change in the dust grain population between the envelope and disc in not yet well characterised, especially in gravitationally unstable systems.

\subsection{Detecting spiral structure}

In order to investigate whether spiral structure can be distinguished across our parameter space, we synthesise observations of our \lime\ images using \casa\  \citep[v4.5.0;][]{McMullin&Waters2007}. We assume our model is located at a distance of 145\,pc in order to examine the observability of self gravitating discs in the nearest star forming regions (e.g. $\rho$ Ophiuchi). We vary the inclination of our disc from 0$^{\circ}$ to 90$^{\circ}$ to cover the range of orientations seen in observations of real systems such as IRAS 16243-2422, which is a protostellar binary with two sources, A and B separated by approximately 600\,au, with near face-on and edge-on disc inclinations respectively. We use two different antenna setups in order to simulate the capabilities of \alma\ Cycle 5 and a fully extended, maximally operational \alma. For \alma\ Cycle 5 we use the antenna configuration that is available and affords us the best angular resolution at each frequency, i.e. 90\,GHz: C43-10, 230\,GHz: C43-10, 300\,GHz: C43-8, 430\,GHz: C43-7, 850\,GHz: C43-7. We simulate observing for 60 minutes using the {\sc{simobserve}} routine in \casa, considering thermal noise with an ambient temperature of 270\,K and a precipitable water vapour given when selecting the `automatic' option in the \alma\ Sensitivity Calculator\footnote{https://almascience.nrao.edu/proposing/observing-tool}, which uses the specified frequency to calculate appropriate values, i.e. 90\,GHz: 5.186\,mm, 230\,GHz: 1.796\,mm, 300\,GHz: 1.796\,mm, 430\,GHz: 0.913\,mm, 850\,GHz: 0.658\,mm. This observation time is expected to be sufficient for detecting the dust continuum emission in the spiral features for the majority of our parameter space. Finally, we use the multiscale {\sc{clean}} algorithm for deconvolution of the synthetic visibility data, in order to effectively recover the extended disc and spiral arm structure.

\smallskip

Here we present only the most detailed image attained when observing our disc model with \alma\ Cycle 5, as seen in Figure \ref{fig:230GHzCycle5}, and compile the results at the remaining frequencies in Appendix \ref{sec:cycle5extra}. The results when using the fully extended, maximally operational antenna configuration are also not displayed here because our overall conclusions are the same, but the images can be found in Appendix \ref{sec:fullalma}. The most detail is recovered at 230\,GHz because there is an optimal combination of angular resolution and sensitivity affording a SNR of approximately 30 within the resolved spirals; we calculate the noise level by taking the RMS of regions within the residual maps free from disc emission or sidelobes. Our observations show that spiral structure is sensitive to the assumed dust opacities, as the contrast between the arm and inter-arm regions is different between the panels, which is important because the state of grain growth is relatively unknown at such early stages of disc evolution. However, as the spiral structure is clearly visible within all panels of Figure \ref{fig:230GHzCycle5}, relatively short observations should be capable of detecting spiral structure in young, gravitationally unstable discs across a wide range of inclinations and possible dust grain opacities.

\smallskip

We note here that is important to cross-check the results of radiative transfer codes in order to ensure there is a sufficient degree of consistency. In order to assess the validity of our results obtained with \lime, we also perform a comparison with synthetic observations produced using \radmc\ with regular gridding, which we detail in \autoref{sec:RADMC3D}. We find that there is a general level of consistency between the flux emitted from the spiral arms, though \radmc\ produces a stronger contrast between arm and inter-arm regions that results in an easier detection of spiral structure within our disc model. Nevertheless, this only serves to strengthen our conclusions that \alma\ should be able to detect spiral structure in a gravitationally unstable disc.

\smallskip

The synthetic observations we present in this paper, including those presented in Appendix \ref{sec:cycle5extra}, \ref{sec:fullalma} and \ref{sec:RADMC3D}, are encouraging for future continuum observations as they show that \alma\ can detect non-axisymmetric structure in young, embedded systems. As aforementioned, this has already been proven with recent observations, however, we can draw further conclusions from our synthetic results in particular. Firstly, unlike real observations to date, we know unequivocally that the spiral features are driven by GIs in our disc model and have successfully demonstrated \alma\ can resolve such GI-driven spirals. Secondly, we only observe our disc for 60 minutes and can already resolve the spiral structure across most of our parameter space at low frequencies. If a longer observational time was adopted, then the spiral features would also become more prominent at higher frequencies, allowing the detection of spirals in a young, gravitationally unstable disc across multiple \alma\ bands. This would allow determinations of the spectral index and give insight into the grain size distribution \citep[e.g.][]{Perez&Carpenter2012}, shedding light on processes such as grain growth and dust trapping \citep[e.g][]{Dipierro&Pinilla2015} in young, embedded discs. Finally, our model is of a protosolar-like disc, which means \alma\ should be able to resolve spiral features in a relatively low-mass disc that is possibly analogous to our early Solar System.

\begin{figure*}
    \includegraphics[width = 0.95\textwidth]{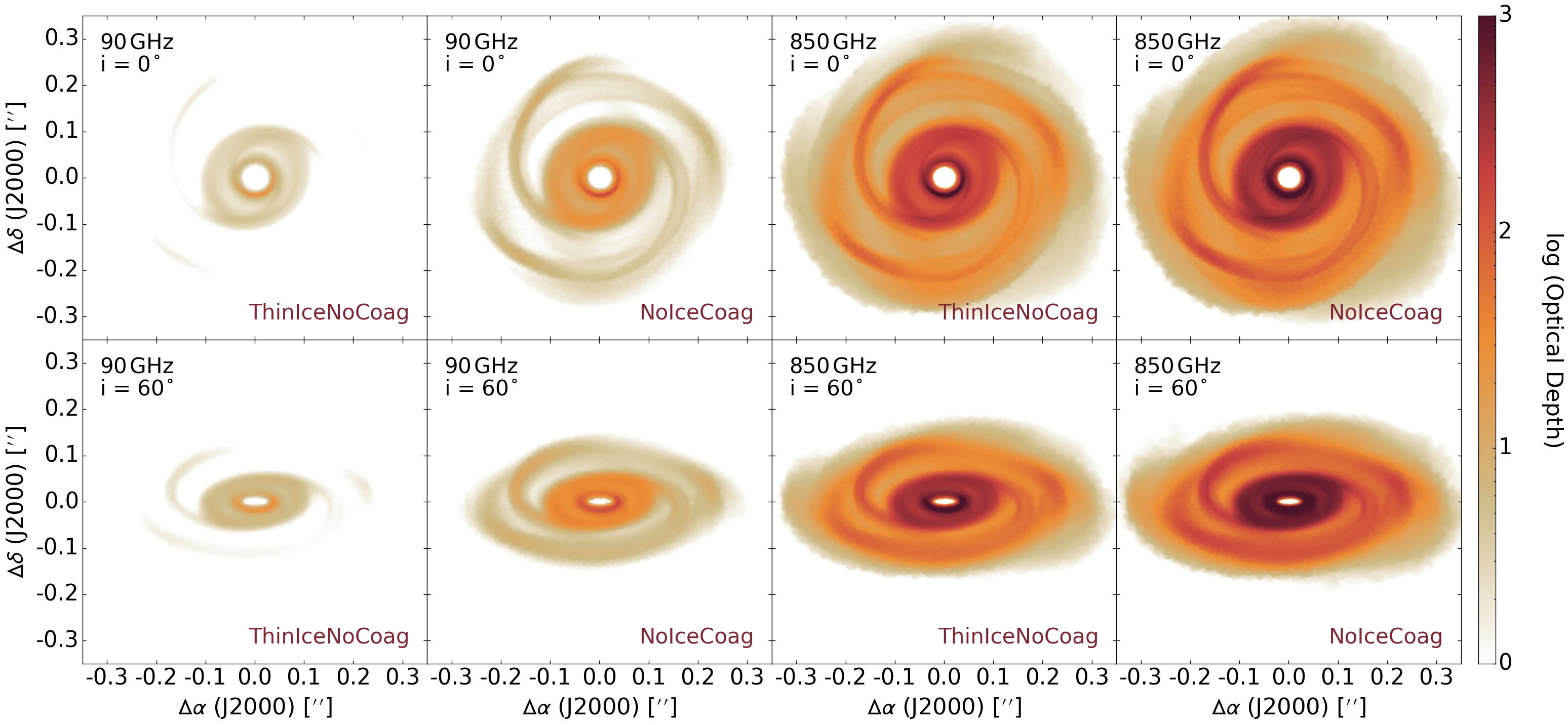}
    \caption{Optical depth along the line-of-sight of our disc model at two different frequencies, inclinations and dust opacity laws.}
    \label{fig:opticaldepths}
\end{figure*}

\subsubsection{Comparison with other studies}

The synthetic observations of our disc model show that \alma\ can detect non-axisymmetric structure in young, embedded systems across a range of frequencies, inclinations and dust opacities. This agrees with \citet{Dipierro&Lodato2014} who use a 3D SPH radiative hydrodynamic code \citep{Lodato&Rice2004} and find that for a gravitationally unstable disc located in the TaurusAuriga and Ophiucus star-forming regions ($\approx$∼140 pc), the spiral structure is readily detectable by ALMA across a range of frequencies, inclinations and disc-to-star mass ratios. However, it should be noted that their model uses a cooling prescription that does not take into account the irradiation emitted by the central star, which implies that the amplitudes of their spiral perturbations are overestimated. Countering this effect somewhat, though, is the fact that \citet{Dipierro&Lodato2014} assume more grain growth has occurred, and hence use larger dust opacities than we consider, which we have shown reduces the detectability of spiral structure. \citet{Dong&Hall2015} use a 3D SPH simulation augmented by a hybrid radiative transfer in order to model both global cooling and radiative transfer \citep{Forgan&Rice2009}, which is comparable to the technique we implement. They find that spiral arms should be detectable in near-infrared scattered light observations of systems with $q$ $\geq$ 0.25, which our disc model is near the threshold of, and also find that their high $q$ models closely resemble real objects, though the validity of GIs operating in these objects is not clear. Our comparison with \citet{Dong&Hall2015} is not like-for-like, as near-infrared scattered light observations only trace the surface layers of discs. However, \citet{Dipierro&Pinilla2015} have shown that GIs can create strong enough surface density perturbations that could be detected in near-infrared scattered light, hence we feel our agreement with \citet{Dong&Hall2015} is noteworthy.

\smallskip

The results of the aforementioned studies, and the results we present in this paper, disagree somewhat with the results of \citet{Hall&Forgan2016}, who use a 1D analytical model \citep{Clarke2009} extended to 2D and 3D, and find spiral structure is only detectable across a narrow parameter space; specifically only at 680\,GHz when adopting a distance of 140\,pc. \citet{Hall&Forgan2016} use a cooling time and $\alpha$ that vary locally so they argue that the relative strengths of the perturbations in their model are much less in the outer part of the disc than they would be in an SPH simulation where $\beta_{cool}$ is fixed at some relatively low value, such as \citet{Dipierro&Lodato2014}. However, this same argument does not hold for \citet{Dong&Hall2015} or our work as non-local global transport is assumed, and hence we argue that GIs should be easier to detect than \citet{Hall&Forgan2016} postulate.

\smallskip

The aforementioned studies in the literature roughly follow the nomenclature that $m$ $\approx$ 1/$q$ \citep[see][Figure 8]{Cossins&Lodato2009}, where $m$ is the number of spiral arms and $q$ is the disc-to-star mass ratio. Moreover, \citet{Dong&Hall2015} conclude that $m$ will be a good diagnostic of the total disc mass. Our disc model does not adhere to this approximation, however, as $m$ = 2 whilst $q$ $\approx$ 0.2, but we emphasise that our disc model is entirely consistent with the long-term behaviour of self-regulated, unstable, radiative discs, particularly considering the highly non-linear behaviour of the system (see the mode analyses in \citet{Mejia&Durisen2005} and \citet{Boley&Durisen2006}, for example, or discussions in \citet{Durisen&Boss2007}). Therefore, whilst $m$ $\approx$ 1/$q$ can be a useful diagnostic, we note that it is not part of a fundamental criterion for gravitationally unstable discs.

\subsection{Observational mass estimates}
\label{sec:massobs}

Disc masses are typically estimated from continuum flux detections using

\begin{equation}
    M_{\rm disc} = \frac{gS_\nu d^2}{\kappa_\nu B_\nu(T)}
    \label{eq:mass}
\end{equation}
where $g$ is the gas-dust mass ratio, usually assumed to be the ISM value of 100, $S_\nu$ is the flux density, $d$ is the distance to the source, $\kappa_\nu$ is an assumed dust opacity and $B_\nu(T_\nu)$ is the Planck function for an assumed dust temperature \citep{Hildebrand1983}. This method assumes that the temperature is uniform and the emission is optically thin. However, young, embedded systems are expected to be optically thick at the wavelengths typically used; for this reason results in the literature usually quote disc masses as lower limits. 

\smallskip

Figure \ref{fig:opticaldepths} shows the optical depth along the line-of-sight through our disc model when observed at the two extreme frequencies. Within this regime of the electromagnetic spectrum the dust opacity increases with frequency, as can be seen in \autoref{tab:freqkappas} and \autoref{fig:opacities}, hence we recover the lowest optical depths at the lowest frequencies. Figure \ref{fig:opticaldepths} clearly demonstrates that the disc spiral features are optically thick across the majority of our parameter space. To clarify, using the lowest dust opacity configuration (ThinIceNoCoag) at 90\,GHz affords an average optical depth of approximately $\tau$ = 1 in the face-on spirals, but this increases with inclination, observing frequency and dust opacity. Note that the opacity of the inner disc, which is where most of the disc mass resides, is even higher than in the spirals. Therefore, young, gravitationally unstable discs are likely to be partially optically thick at millimetre wavelengths and substantially optically thick at submillimetre wavelengths, which could result in significantly underestimated disc masses. As we know the exact mass of our disc model we can explore how inaccurate such disc mass estimates can be.

\smallskip

We assume two different dust temperatures, $T_ {\rm dust}$ = 20\,K and 40\,K when calculating the disc mass. The former is chosen to mimic the calculations performed using observations of more evolved, Class II \citep[e.g][]{Ansdell&Williams2016}, based on the median temperature of Class II discs in Taurus-Auriga \citep{Andrews&Williams2005}. However, as GIs can heat young discs significantly (see \citetalias{Evans&Ilee2015}), we also adopt a higher dust temperature, $T_{\rm dust}$ = 40\,K. Coincidentally, this is the dust temperature \citet{Tobin&Kratter2016} use when deriving the mass of the recently formed fragment in L1448 IRS3B, as they argue the region around the fragment should be warmed by turbulence. Moreover, this is the peak brightness temperature, $T_B$, towards this fragment, and in optically thick regions $T_B$ $\approx$ $T_ {\rm dust}$.

\begin{figure*}
    \includegraphics[width = 0.95\textwidth]{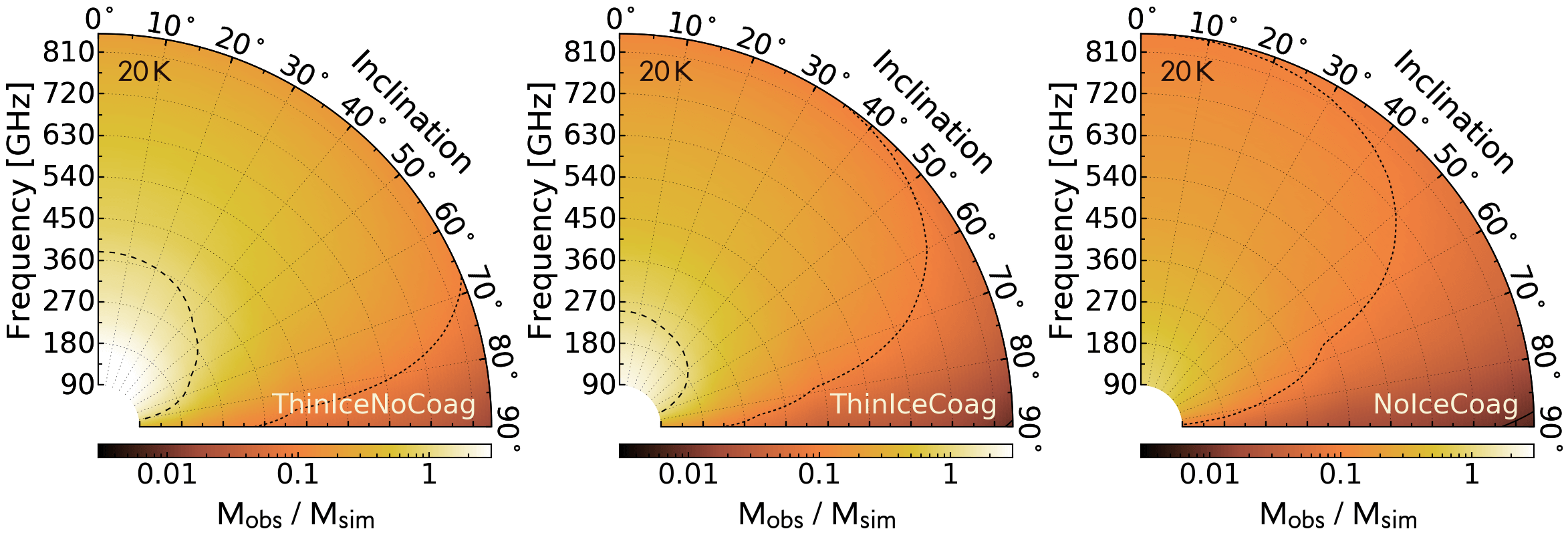}
    \caption{Fraction of actual disc mass derived from observations at various inclinations and frequencies, for three different dust grain properties (see Table \ref{tab:freqkappas}), assuming a constant dust temperature of 20\,K. The contours denote fractions of 0.01 (solid line), 0.1 and 1.0 (longest-dashes line). Note that the face-on disc is inclined at 0$^\circ$.}
    \label{fig:20Kmass}
\end{figure*}

\begin{figure*}
    \includegraphics[width = 0.95\textwidth]{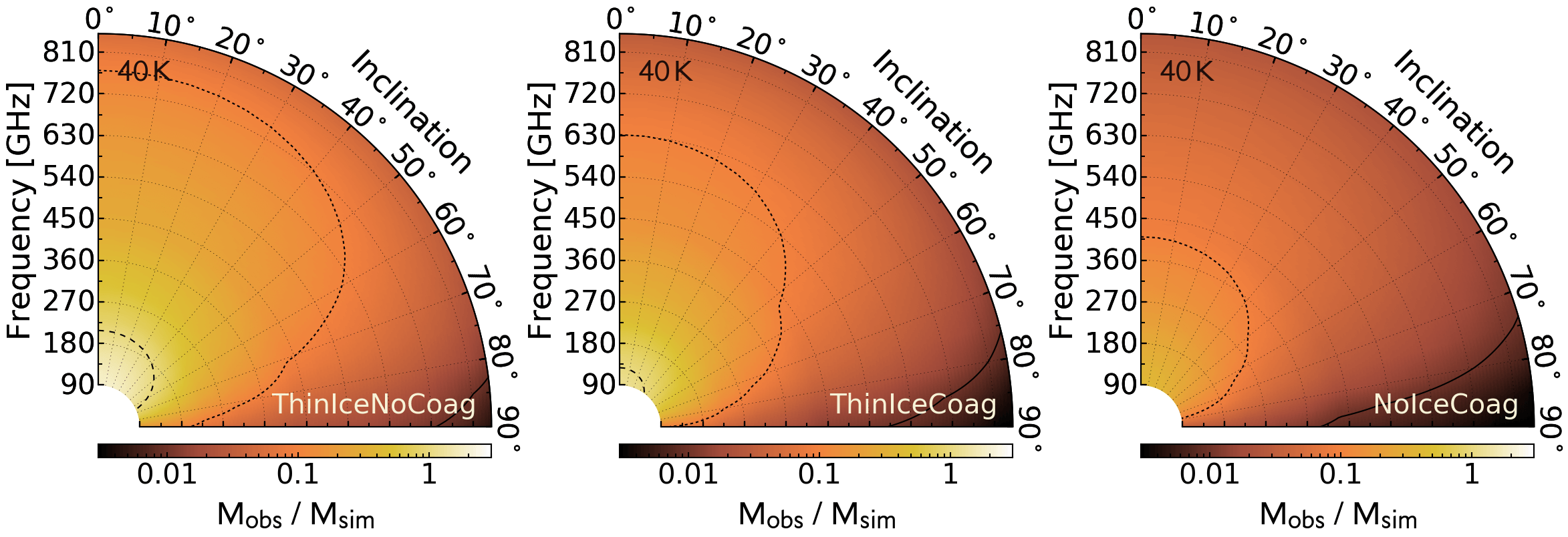}
    \caption{Same as Figure \ref{fig:20Kmass} but assuming a constant dust temperature of 40\,K.}
    \label{fig:40Kmass}
\end{figure*}

We estimate the flux density from the Cycle 5 \alma\ synthetic observations, shown in Figures \ref{fig:90GHzCycle5}-\ref{fig:850GHzCycle5}, by fitting regions to the disc emission that varied with inclination and frequency. We then calculate disc masses across the entire parameter space for our observations produced using both the Cycle 5 antenna configuration and the fully extended, maximally operational antenna configuration. However, as there are minimal differences between the masses derived from the two antenna setups, we omit the fully operational \alma\ results.
\smallskip

Figures \ref{fig:20Kmass} and \ref{fig:40Kmass} show how the ratio of the observationally derived mass, $M_{\rm obs}$, to the actual mass of our simulated disc model, $M_{\rm sim}$, varies as a function of frequency, inclination, dust opacity and dust temperature. As the blackbody flux, $B_\nu$, increases with temperature, we recover a lower disc mass at a higher dust temperature. We also recover a lower disc mass if we use a dust grain configuration that results in larger optical depths (see Table \ref{tab:freqkappas}), as we are observing emission from a smaller fraction of the disc. Similarly, if we observe at higher frequencies then we are not able to peer as far through the disc, as can be seen in Figure \ref{fig:opticaldepths}, so we also recover a lower disc mass. Finally, as the inclination of the disc increases, the recovered mass decreases because the morphology of the disc loses distinction and the line-of-sight column density increases. As a result, the recovered flux density, $S_\nu$, decreases, particularly as the orientation approaches edge-on.

\smallskip

In summary of these results, the derived disc mass is sensitive to the disc properties as well as the observational setup. However, some combinations of parameters are probably unrealistic for young, gravitationally unstable discs. For instance, when adopting low dust opacities (ThinIceNoCoag), which assumes no coagulation has occurred at such early stages of disc evolution, and a low dust temperature, $T_ {\rm dust}$ = 20\,K, the derived disc mass can be significantly larger than the actual mass at low frequencies. Indeed, observing our disc model close to face-on at 90\,GHz results in an observed mass, $M_{\rm obs}$ = 4$M_{\rm sim}$ when adopting these parameter values. These overestimates occur for two primary reasons. Firstly, using an underestimated dust temperature results in a lower Planck flux when far from the Rayleigh-Jeans limit, which increases the disc mass. As aforementioned, $T_ {\rm dust}$ = 20\,K is the median temperature of Class II discs and therefore, due to the shock-heating of GIs, is likely an underestimate for young, gravitationally unstable discs. Secondly, this dust configuration (ThinIceNoCoag) disagrees with the underlying dust distribution in the hydrodynamic model across our observational frequency range (see \autoref{fig:opacities}), which assumes dust grains up to 1\,mm are present. As observations indicate that larger grains are already present in the envelopes of Class I objects \citep[e.g.][]{Miotello&Testi2014}, and are most likely explained by grain growth within the disc \citep{Wong&Hirashita2016}, low dust opacities are likely inaccurate in young, gravitationally unstable discs. Note that whilst assuming grain growth perhaps contradicts our assumption of well mixed dust and gas, we assume there has not been sufficient time for the dust and gas to become decoupled because our disc model is very turbulent.

\smallskip

Now, as we have argued that a higher dust opacity and dust temperature are most appropriate for our disc model, and we have shown that `NoIceCoag' is self-consistent with the disc simulation, we focus on the right panel of Figure \ref{fig:40Kmass}. We recover maximum disc masses when considering a face-on disc, which results in: $M_{\rm obs}$ = 0.41$M_ {\rm sim}$ at 90\,GHz; $M_{\rm obs}$ = 0.21$M_ {\rm sim}$ at 230\,GHz; $M_ {\rm obs}$ = 0.15$M_ {\rm sim}$ at 300\,GHz; $M_ {\rm obs}$ = 0.09$M_ {\rm sim}$ at 430\,GHz; and $M_ {\rm obs}$ = 0.03$M_ {\rm sim}$ at 850\,GHz. Deriving disc masses is therefore preferable at lower frequencies. However, as angular resolution is inversely proportional to frequency, obtaining accurate masses and spatially resolving spiral structure are likely mutually exclusive objectives currently. Overall, when using what we understand to be realistic parameters, protoplanetary disc masses derived from continuum observations in the millimetre regime could underestimate the actual mass in a face-on disc by a factor of 2.5--30, depending on observing frequency. Moreover, this underestimate worsens as the disc inclination increases.

\begin{figure*}
    \includegraphics[width = 0.95\textwidth]{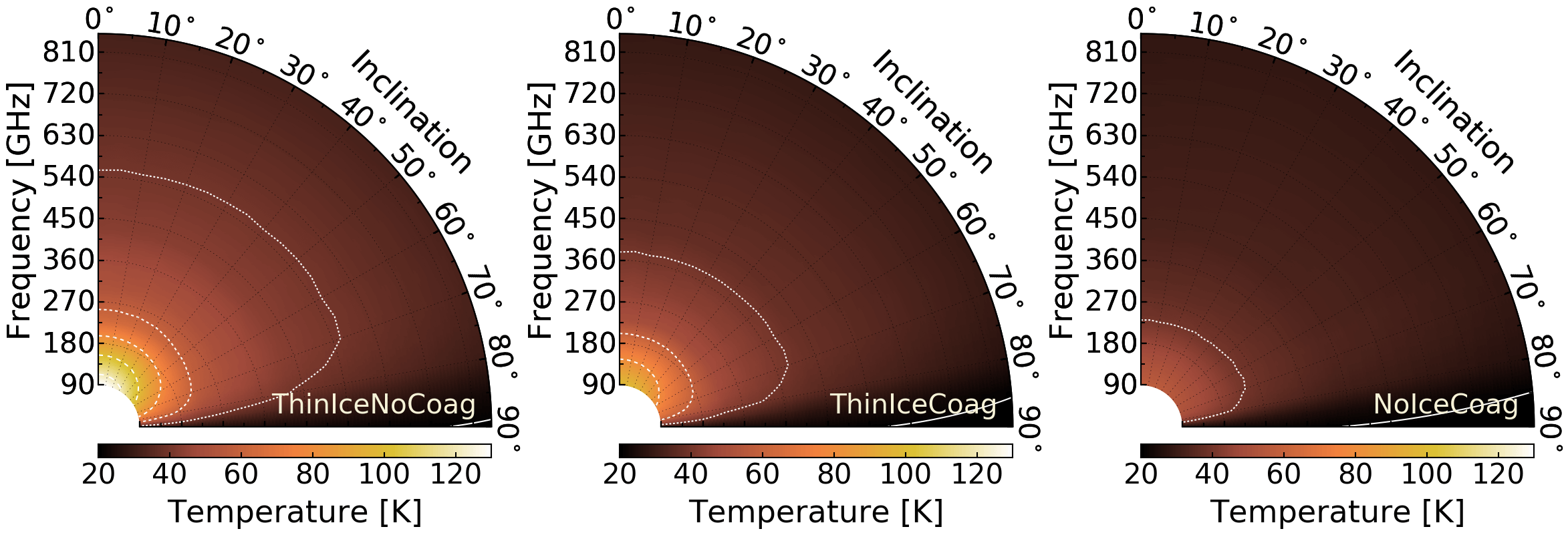}
    \caption{Mass-weighted average temperature of the optically thin emission at various inclinations and frequencies, for three different dust grain properties (see Table \ref{tab:freqkappas}). The contours denote temperatures from 20\,K (solid line) to 120\,K (longest-dashes line) in 20\,K increments.}
    \label{fig:accuratetemps}
\end{figure*}

\subsubsection{Comparison with other studies}

Our mass underestimation results are consistent with the predictions of \citet{Forgan&Rice2013}, who use self-gravitating disc models to match observations of the Class 0 protostar L1527 IRS, and continuum results of \citet{Douglas&Caselli2013}, who simulate a GI-driven disc representing a violent phase of evolution, with a disc-to-star mass ratio nearly double what we consider. This result also appears consistent with the results of \citet{Dunham&Vorobyov2014}, who report a mass underestimate in a gravitationally unstable disc similar to our model, of a factor of 2--3 at millimetre wavelengths and a factor of 10 or more at submillimetre. \citet{Dunham&Vorobyov2014} calculate disc masses assuming a source distance of 250\,pc, and so, as we use a source distance of 145\,pc, we should perhaps expect less extreme underestimates for our disc model based on their results. However, this is not the case and one possible explanation for this is that the radiative transfer calculations performed by \citet{Dunham&Vorobyov2014} are 2-dimensional and do not include the full, non-axisymmetric structure of the disc. As a result, spiral features, which we find to be extremely optically thick at most wavelengths (see \autoref{fig:opticaldepths}) are not represented accurately. Despite this, we draw the same conclusions as 
\citet{Dunham&Vorobyov2014} that masses derived from observations of young, embedded discs could be significantly underestimated, except we find that these underestimates may be even more pronounced than they predicted. If this is the case it could have important effects on the derived properties and classifications of young systems. In particular, discs that have been observed and determined not to contain enough mass to be gravitationally unstable could actually be GI-driven.

\subsubsection{Accurate dust temperatures}
\label{sec:accuratetemps}

It is important to acknowledge that the assumed dust temperature in \autoref{eq:mass} is an estimated average of the optically thin emission temperature, which is typically not well constrained in young, embedded systems. Therefore, it is extremely worthwhile to assess the validity of the dust temperatures we have used in deriving the mass of our disc model thus far, and compare with typical assumptions made when deriving masses from observations.

\smallskip

The actual dust temperature of the optically thin emission varies as a function of the optical depth. Hence, the appropriate average dust temperature to assume when using \autoref{eq:mass} will change as a function of frequency, inclination and dust opacity. We explore this effect in our disc model by calculating the mass-weighted average temperature of the emission above the $\tau$ = 1 surface, which characterises the optically thin emission, in a similar method as described in Section \ref{sec:tausurface}. Figure \ref{fig:accuratetemps} demonstrates these results for our disc model and, as can be seen, for all dust opacities we consider, the mass-weighted average temperature at 90\,GHz is higher than at shorter wavelengths. This occurs due to the increase in optical depth with frequency, and this effect is much more pronounced when assuming low dust opacities (ThinIceNoCoag) compared to high dust opacities (NoIceCoag). At 270\,GHz and beyond, it becomes reasonable to adopt a constant dust temperature, assuming $i$ \textless\ 80$^\circ$. At near-edge-on inclinations the morphology of the disc is obscured and the mass-weighted average temperature approaches the minimum value we recover, $T_ {\rm dust} \approx$ 20\,K. 

\smallskip

Beyond Band 3 of \alma\ (\textgreater 116\,GHz), our results indicate that the average dust temperature is $T_ {\rm dust} \approx$ 30--40\,K in our GI-driven, embedded protoplanetary disc, except when adopting low dust opacities (ThinIceNoCoag). If observations are performed at Band 3 (84--116\,GHz), where the optical depth is lowest, the assumed dust temperature should be increased significantly as more of the shock-heated regions can be observed; for our disc model, $T_ {\rm dust} \approx$ 55--120\,K at 90\,GHz, depending on the dust opacity law adopted. Our findings, which are representative of a late Class I/early Class 0 system, disagree with the recommendation of \citet{Dunham&Vorobyov2014} to adopt $T_ {\rm dust} \approx$ 30\,K in a Class 0 source and $T_ {\rm dust} \approx$ 15\,K in a Class I source. This is most likely because, as aforementioned, their radiative transfer modelling neglects the shock heating driven by GIs, and so their temperature estimates of the optically thin emission could be underestimated. This would then mean that their derived masses are likely to be overestimated. 

\smallskip

Overall, the method for deriving masses from continuum flux observations is questionable for two primary reasons. Firstly, the dust temperatures in a gravitationally unstable disc will cover a large range (approximately 20--250\,K in our disc model), and reducing this to a mass-weighted average dust temperature neglects the complex non-axisymmetric structure of the disc. Secondly, the dust temperature is an uncertain property as it is dependent on the dust opacity, which itself is poorly constrained due to the unconfirmed state of grain growth in young, embedded systems. Moreover, using a dust temperature that is inappropriately low for the system in question can inadvertently compensate for the large optical depths caused by large dust opacities. For example, we can recover $M_ {\rm obs}$ $\approx$ $M_ {\rm sim}$ for our disc model when adopting large opacities (NoIceCoag) if we use $T_ {\rm dust}$ = 20\,K, which as we have shown is an unrealistic dust temperature for our disc model. Therefore, even in this case, where the observational mass approaches the actual mass, the calculated disc mass is not accurate. Hence, mass estimates derived from continuum flux observations of embedded, gravitationally unstable discs are inherently unreliable.

\section{Conclusions}
\label{sec:conclusions}

\subsection{\lime\ optimisations}

We have produced continuum emission maps of a protoplanetary disc model using \lime\ and in Section \ref{sec:LIME} have explored in-depth how accurate images can be attained. The main results from this body of work, which focuses on continuum images, are as follows:

\begin{itemize}[leftmargin=0.0cm, itemindent=1.0cm]

\item The optimal grid point weighting function is dependent on the model used and features of interest within the model. We use number density weighting, which is the vanilla \lime\ default, and find that $n_0$ = 1$\times10^{12}\mathrm{cm^{-3}}$ and $w$ = 0.5 highlight the spiral features within our disc model.

\item A convergence test should be used to find the number of grid points necessary to produce an accurate continuum emission image. This will be dependent on the model, but for our young, gravitationally unstable protoplanetary disc model we find that a minimum of $2\times10^5$ grid points should be used. Note, however, that more grid points is always preferable if the computational cost can be afforded. This strategy should also be applied to line images.

\item The contribution to emission intensity in high optical regions is negligible. Grid points can therefore be omitted where $\tau$ \textgreater\ 3, which, as these are the densest regions, increases the number of grid points in the observed regions significantly. As a result, this method results in an improvement in the accuracy of the continuum images for our disc model when compared to vanilla \lime\ images produced using the same number of grid points. Hence, these findings demonstrate that specialised grid construction can improve the efficiency and accuracy of \lime\ images.

\item Averaging multiple \lime\ runs using identical inputs improves the accuracy of the continuum emission image because the grid point coverage is effectively increased. As multiple instances of \lime\ can be run simultaneously, we see no strong argument against averaging consistent \lime\ images, barring strict computational or time restraints.

\end{itemize}

Although \lime\ is specialised for line emission, we have presented our findings as an indicator that care should be taken even when using a radiative transfer code to produce continuum emission images, in order to ensure these images are accurate. When considering line emission images, the issues we have discussed in this paper are amplified due to the complexity of the radiative transfer equation for molecular line transitions. Therefore, even more care should be taken in this scenario. Our results regarding the continuum should be applicable to molecular line images, which we will explore in a forthcoming publication.

\smallskip

\subsection{Continuum emission and mass estimates}

In Section \ref{sec:LIME} we produced synthetic observations of our disc model using \casa, across a large parameter space consisting of different frequencies, inclinations and dust opacities. Our model represents a $0.17\,\mathrm{M}_{\odot}$ protoplanetary disc that surrounds a $0.8\,\mathrm{M}_{\odot}$ protostar likely to evolve into a Solar-like star. As a result our work may be indicative of observations of an object similar to our early Solar System. The main conclusions we have drawn from these synthetic observations are as follows:

\begin{itemize}[leftmargin=0.0cm, itemindent=1.0cm]

\item Using the antenna configuration for \alma\ Cycle 5 and observing for 60 minutes reveals that, at 90--300\,GHz, spiral features can be readily detected in a 0--60$^\circ$ inclined young, gravitationally unstable disc at a distance of 145\,pc. Spiral features can also be distinguished at near-face-on discs at 430\,GHz, but can only be identified at 850\,GHz when the dust opacity is very low. For Cycle 5, we find that the optimal frequency to observe non-axisymmetric structure in embedded objects is 230\,GHz (1.3\,mm).

\item Using a fully extended, maximally operational \alma\ affords similar results to Cycle 5 observations at lower frequencies as the angular resolution and sensitivity do not change significantly. However, at 430\,GHz and 850\,GHz the angular resolution is much improved; we obtain a minimum spatial resolution of 1.6\,au and 0.8\,au at 430\,GHz and 850\,GHz, respectively. Unfortunately, the spirals are still not distinguishable at these higher angular resolutions because the noise level increases with frequency. Nevertheless, the non-axisymmetric inner disc structure can be seen in extraordinary detail, particularly in inclined discs. Hence, once \alma\ reaches maximum capacity, observations in Bands 8, 9 and 10 could be used to probe the innermost regions of embedded discs and potentially offer unprecedented insight into their evolution mechanisms.

\item We find that the majority of our gravitationally unstable disc model is substantially optically thick at frequencies beyond 90\,GHz. This is because the spiral waves and inner disc contain a considerable amount of mass in vertically elongated regions, which results in large line-of-sight optical depths. As a result, observations of embedded discs should be performed at long wavelengths where the optical depth is lowest.

\item As the GIs within a gravitationally unstable disc heat the disc material on a global scale, using a dust temperature of 20\,K, which is typically used for more evolved discs, is likely inaccurate. Instead, we find that the mass-weighted average temperature of the optically thin regions within our disc model is $T_ {\rm dust} \approx$ 30--40\,K at high frequencies ($\geq$ 230\,GHz). At low frequencies (90\,GHz), $T_ {\rm dust} \approx$ 55--90\,K as the disc is less optically thick, which means more of the shock-heated spiral structure is observed.

\item Using the flux density of observations to estimate the mass of young, embedded discs is an inherently flawed procedure for two primary reasons. Firstly, condensing the 3-dimensional spiral structure into one value for the dust temperature and one value for the dust opacity across the entire disc is misrepresentative. Secondly, the dust temperature and dust opacity in optically thick objects are largely unknown quantities, hence mass derivations are unreliable.

\item Assuming what is believed to be an appropriate dust temperature and opacity within a young, gravitationally unstable disc, the disc mass derived from observations can be underestimated by a a factor of 2.5--30 in a face-on disc, depending on frequency. Moreover, this underestimate worsens as the disc inclination increases. If our assumptions are appropriate, then this could retrospectively validate GIs in discs previously thought not massive enough to be gravitationally unstable, which could have a significant impact on the understanding of the formation and evolution of protoplanetary discs.

\end{itemize}

The results of our synthetic observations are incredibly encouraging for the future and suggest that spiral features in young, embedded, gravitationally unstable discs should be detectable by recent and future cycles of \alma\ at millimetre wavelengths, and perhaps at submillimetre wavelengths given optimal conditions and long observation times. Our synthetic observations support the recent detection of spiral structure in protoplanetary discs \citep[e.g.][]{Tobin&Kratter2016, Perez&Carpenter2016}, which strengthens the argument that these non-axisymmetric features are driven by gravitational instabilities. However, as these objects are more complex than our disc model, this argument is still tentative. Therefore, more observations are needed to assess the validity of GIs in young, embedded objects. If such observations are performed across multiple wavelengths, we should be able to attain unprecedented levels of detail of the disc morphology, which will allow the importance of gravitational instabilities on the formation and evolution of protoplanetary discs to be understood. 

\smallskip

The synthetic observations of our model disc suggest that the mass of a young, gravitationally unstable disc cannot be accurately constrained as these objects are likely optically thick across a broad range of \alma\ frequencies. Therefore alternative approaches should be adopted, such as extracting kinematic information from molecular line observations. Moreover, when considering line emission, observing spiral structure in certain, optically thin, molecular tracers could suggest that localised heating of the disc material is occurring \citepalias[see][]{Evans&Ilee2015}, which would strengthen the argument for GI-driven spiral structure. Therefore, in the near future, we will use \lime\ and our gridding optimisations to explore the ability of \alma\ to extract kinematic information and resolve spiral structure in commonly observed molecular line transitions, in order to quantify the accuracy of disc masses derived from line images and to support the findings of our synthetic continuum observations.

\section{Acknowledgements}
\label{sec:acknowledgements}

We would like to thank the anonymous referee for constructive comments that have improved the clarity of this manuscript. MGE gratefully acknowledges a studentship from the European Research Council (ERC; project PALs 320620).  JDI gratefully acknowledges support from the DISCSIM project, grant agreement 341137, funded by the European Research Council under ERC-2013-ADG.  TWH, PC and LSz acknowledge the financial support of the European Research Council (ERC; project PALs 320620).  ACB's contribution was supported, in part, by The University of British Columbia and the Canada Research Chairs program.

\bibliographystyle{PaperII}
\bibliography{library}

\appendix
\section[Cycle5]{Cycle 5 continuum emission images at 90, 300, 430 and 850\,GHz} 
\label{sec:cycle5extra}

\begin{figure*}
    \includegraphics[width = 0.95\textwidth]{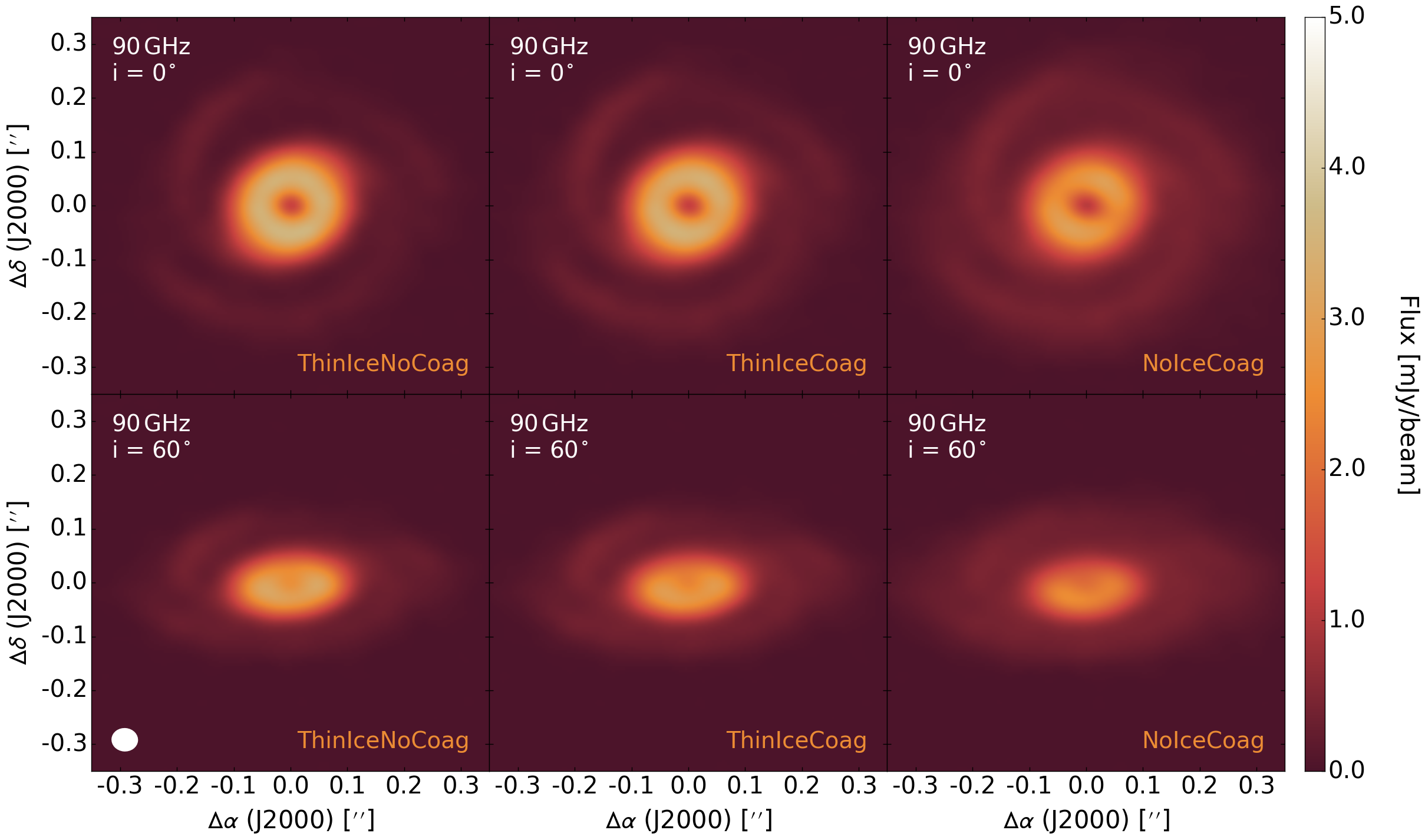}
    \caption{Continuum emission images of our disc model at 90\,GHz for three different dust configurations and two different inclinations, synthesised using the \alma\ Cycle 5 antenna configuration. The white ellipse in the lower left indicates the size of the beam, which is 0.046$\,\times\,$0.044\,arcsec and is constant across all panels. The noise is approximately 11\,\textmu Jy per beam across the parameter space.}
    \label{fig:90GHzCycle5}
\end{figure*}

\begin{figure*}
    \includegraphics[width = 0.95\textwidth]{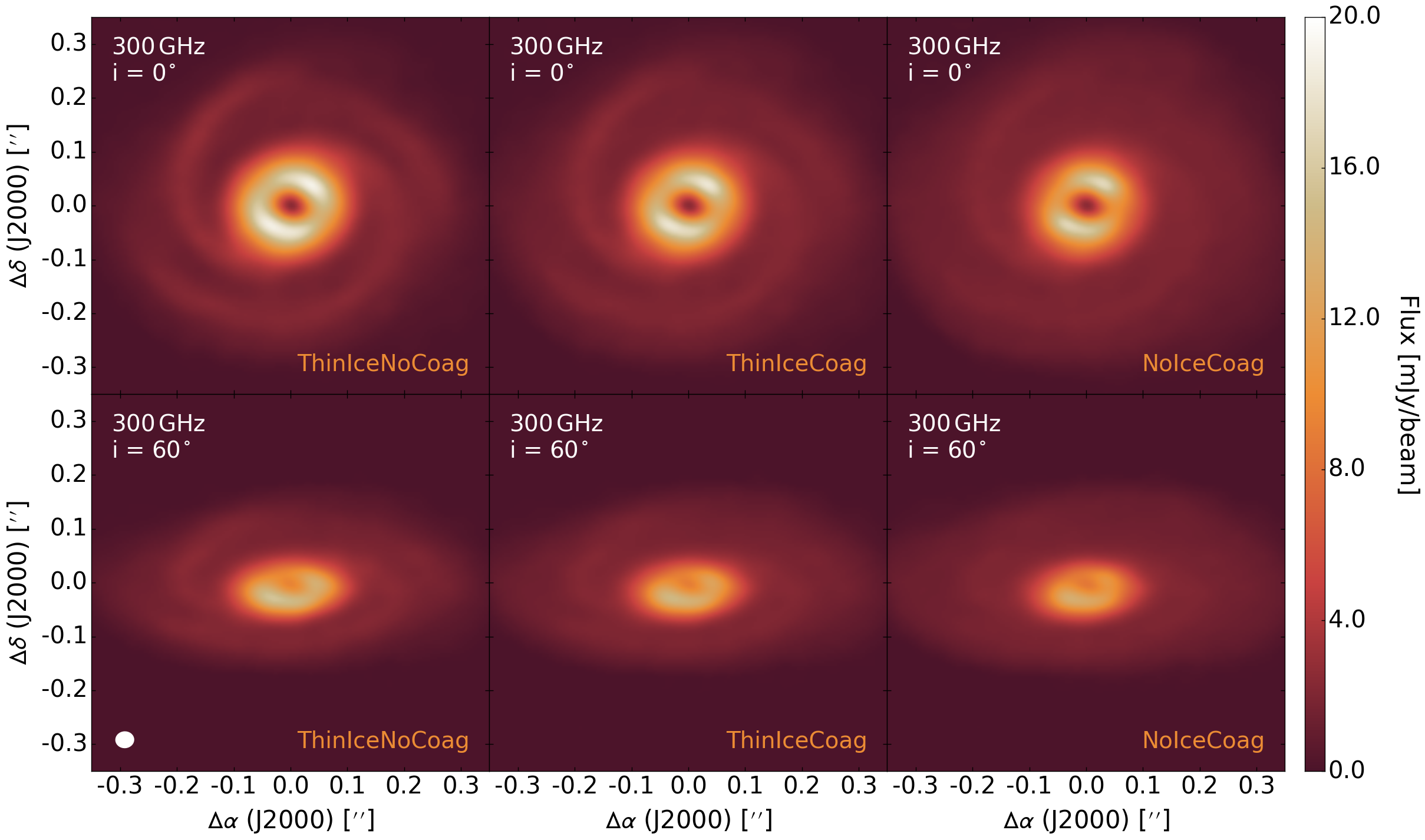}
    \caption{Same as Figure \ref{fig:90GHzCycle5} but synthesised at 300\,GHz with a beamsize of 0.033$\,\times\,$0.031\,arcsec and a noise level of approximately 26\,\textmu Jy per beam.}
    \label{fig:300GHzCycle5}
\end{figure*}

\begin{figure*}
    \includegraphics[width = 0.95\textwidth]{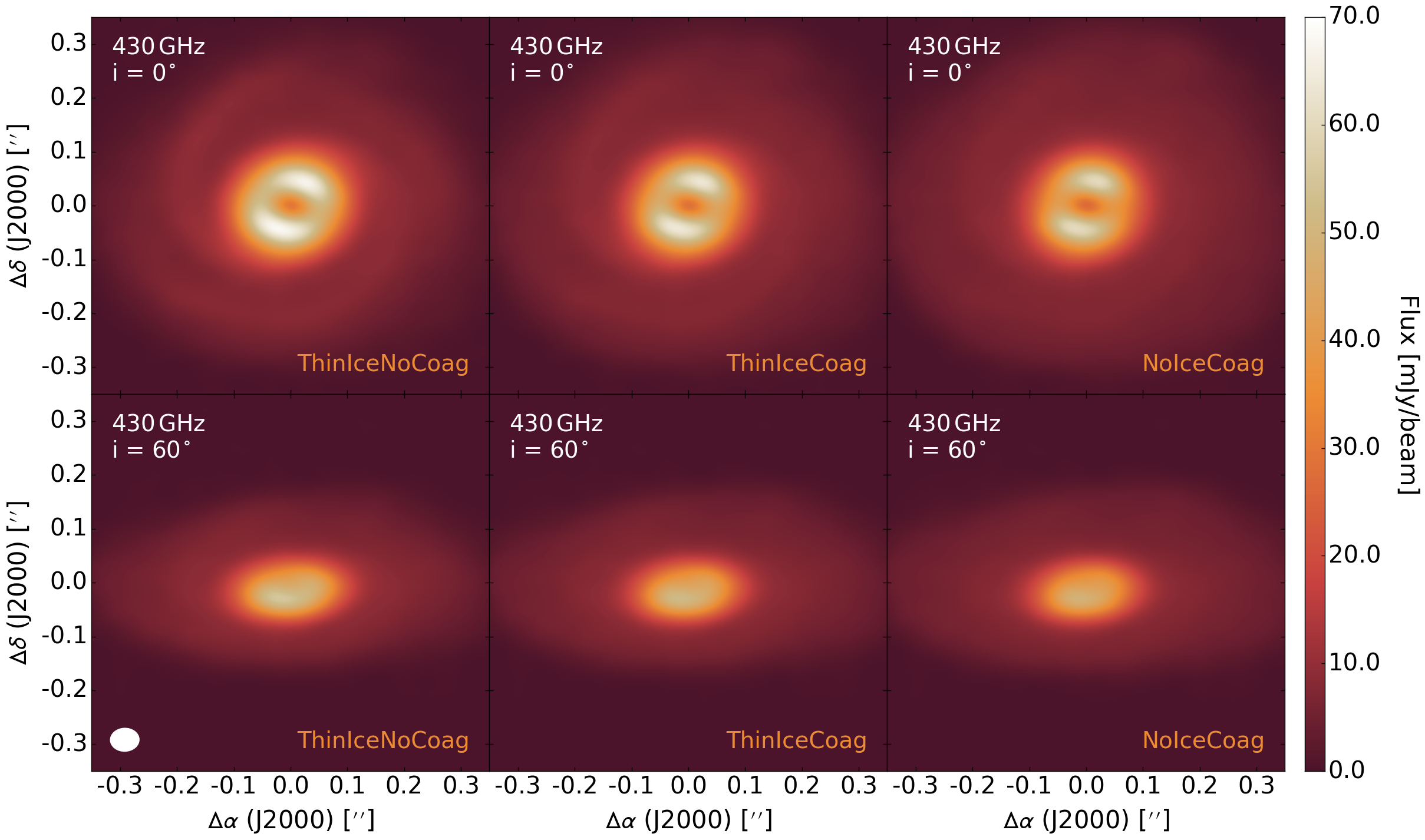}
    \caption{Same as Figure \ref{fig:90GHzCycle5} but synthesised at 430\,GHz with a beamsize of 0.052$\,\times\,$0.045\,arcsec and a noise level of approximately 115\,\textmu Jy per beam.}
    \label{fig:430GHzCycle5}
\end{figure*}

\begin{figure*}
    \includegraphics[width = 0.95\textwidth]{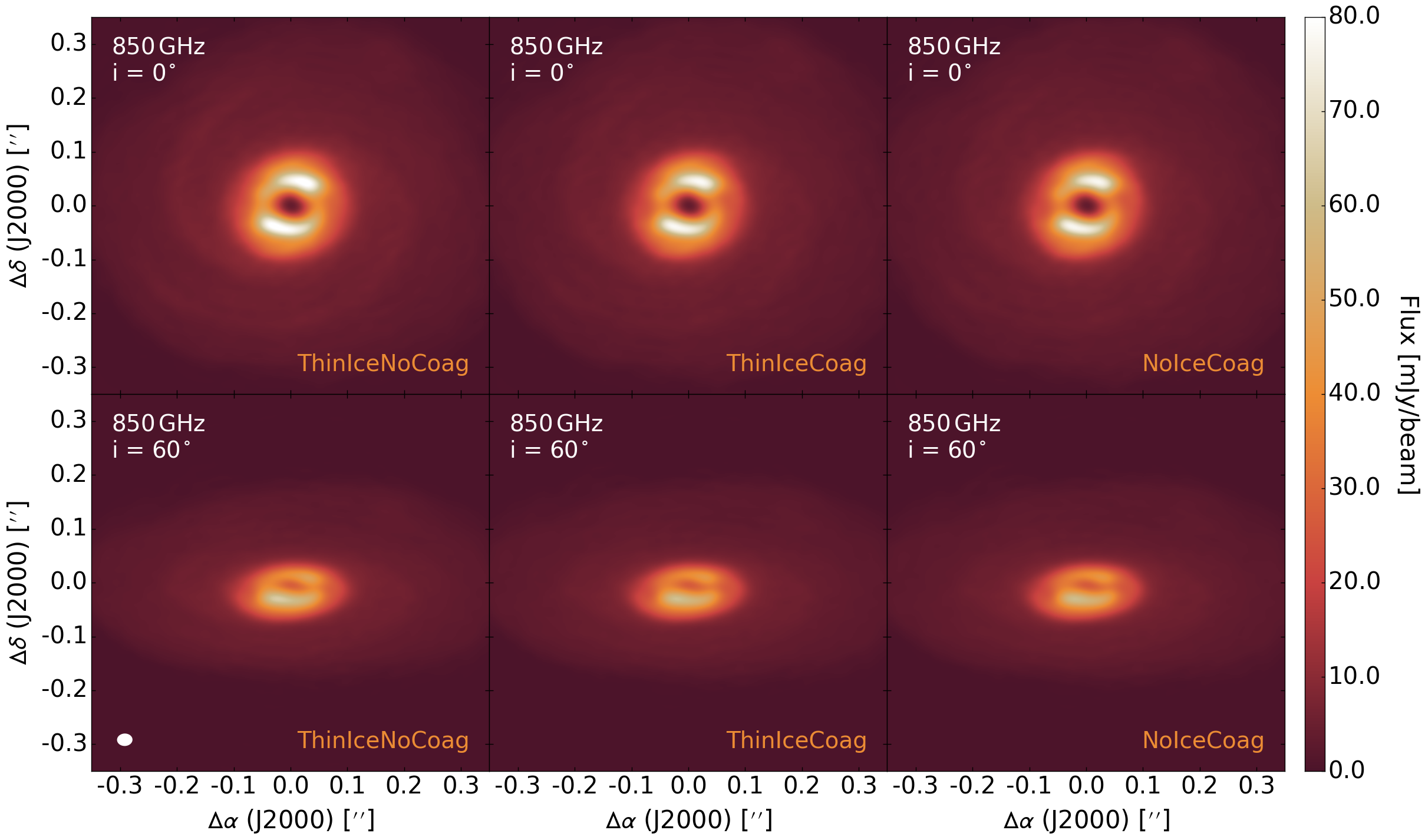}
    \caption{Same as Figure \ref{fig:90GHzCycle5} but synthesised at 850\,GHz with a beamsize of 0.027$\,\times\,$0.023\,arcsec and a noise level of approximately 110\,\textmu Jy per beam.}
    \label{fig:850GHzCycle5}
\end{figure*}

Figures \ref{fig:90GHzCycle5}-\ref{fig:850GHzCycle5} represent the capabilities of \alma\ Cycle 5 when using our disc model and observational setup, with an integration time of 1 hour. We use the antenna configuration that is available and affords us the best angular resolution at each frequency, i.e. 90\,GHz: C43-10, 300\,GHz: C43-8, 430\,GHz: C43-7, 850\,GHz: C43-7. As can be seen, the spiral structure is distinguishable in a face-on disc across all frequencies and dust opacities. At an inclination of 60\,$^\circ$ the spirals are only identifiable at 300\,GHz or lower, and only convincingly when assuming a lower dust opacity.

\section[fullalma]{Fully operational \alma\ continuum emission images at 300, 430 and 850\,GHz} 
\label{sec:fullalma}

\begin{figure*}
    \includegraphics[width = 0.95\textwidth]{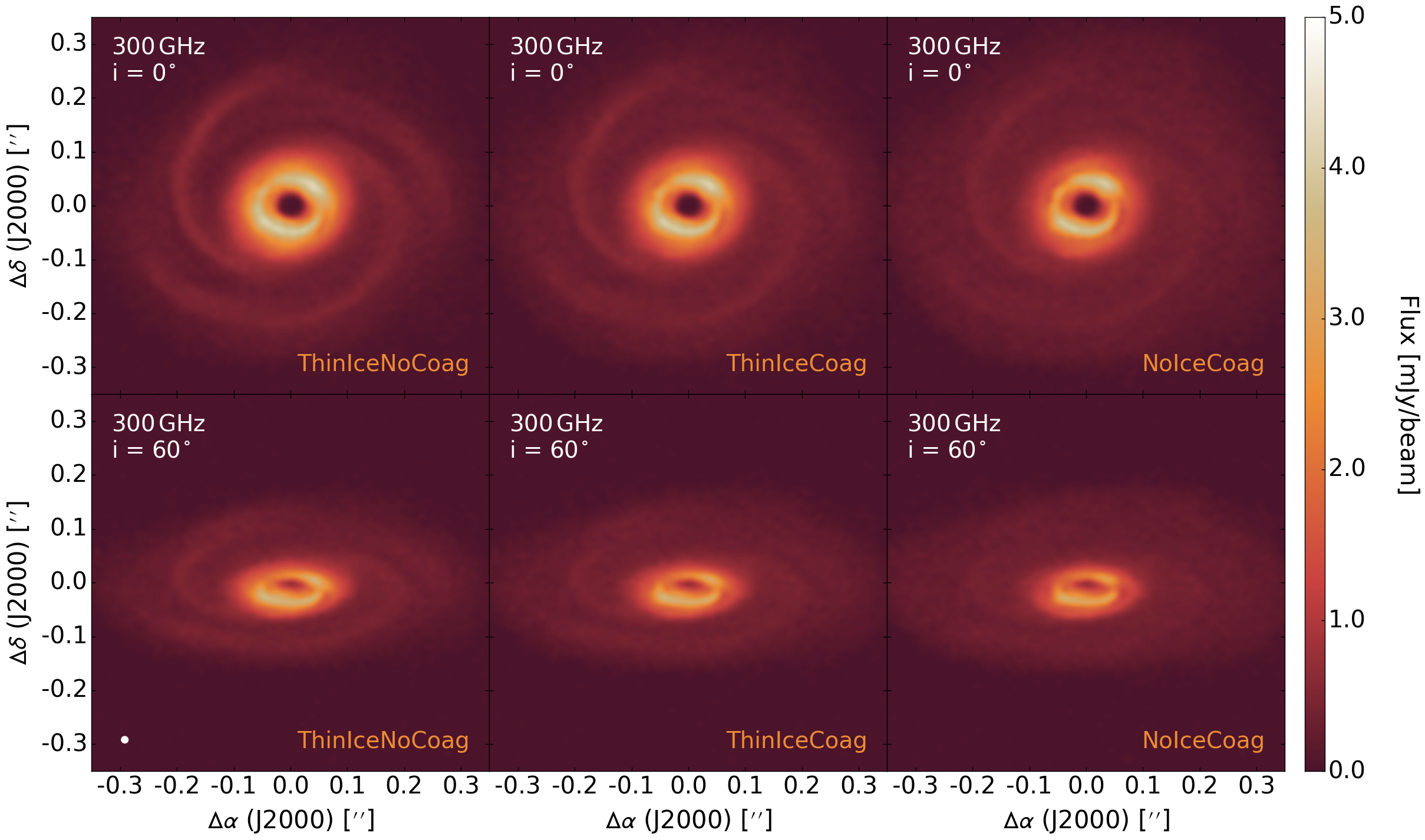}
    \caption{Continuum emission images of our disc model at 300\,GHz for three different dust configurations and two different inclinations, synthesised using a fully extended, maximally operational \alma\ antenna configuration. The white ellipse in the lower left indicates the size of the beam, which is 0.014$\,\times\,$0.014\,arcsec and is constant across all panels. The noise is approximately 20\,\textmu Jy per beam across the parameter space.}
    \label{fig:300GHzFullALMA}
\end{figure*}

\begin{figure*}
    \includegraphics[width = 0.95\textwidth]{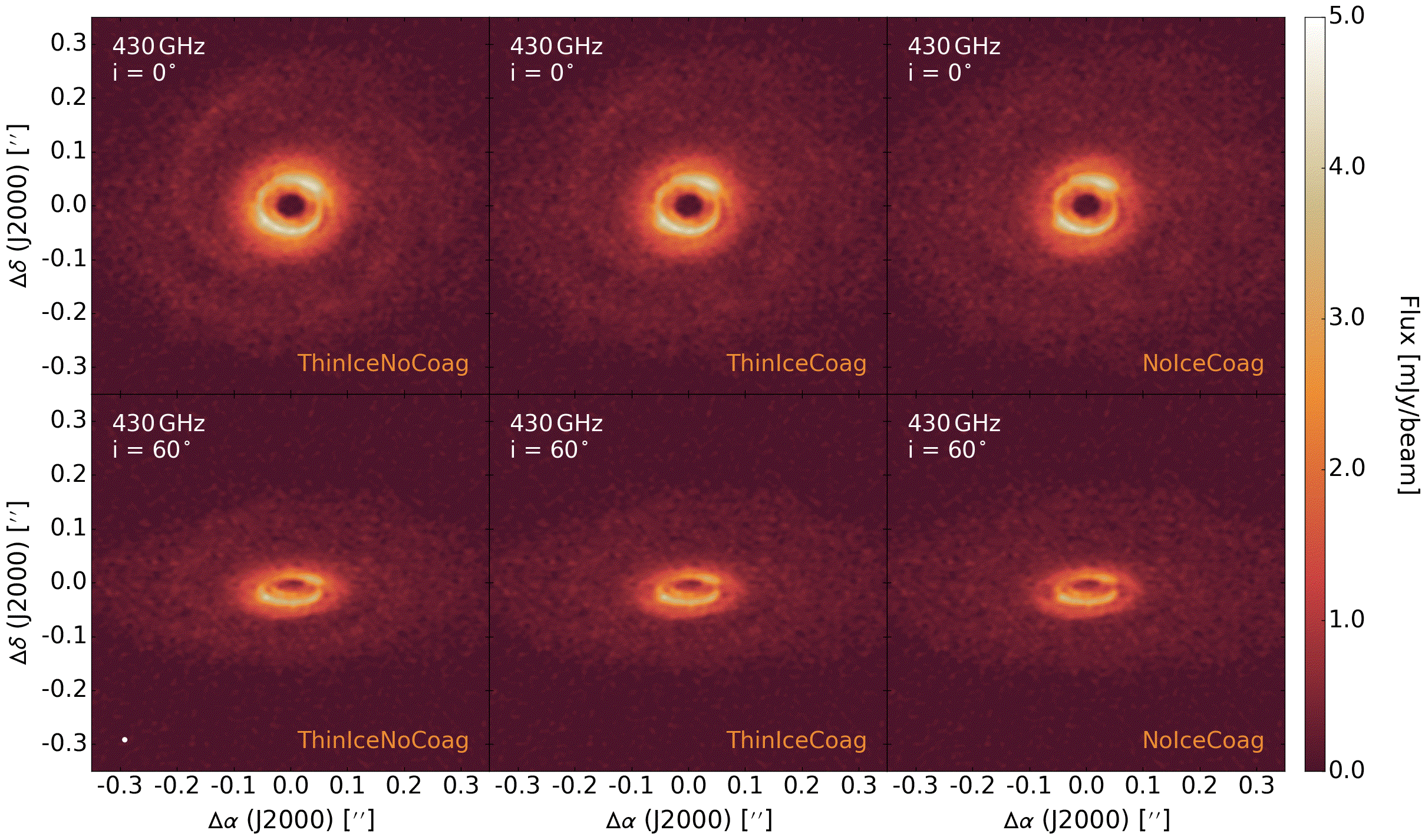}
    \caption{Same as Figure \ref{fig:300GHzFullALMA} but synthesised at 430\,GHz with a beamsize of 0.010$\,\times\,$0.010\,arcsec and a noise level of approximately 85\,\textmu Jy per beam.}
    \label{fig:430GHzFullALMA}
\end{figure*}

\begin{figure*}
    \includegraphics[width = 0.95\textwidth]{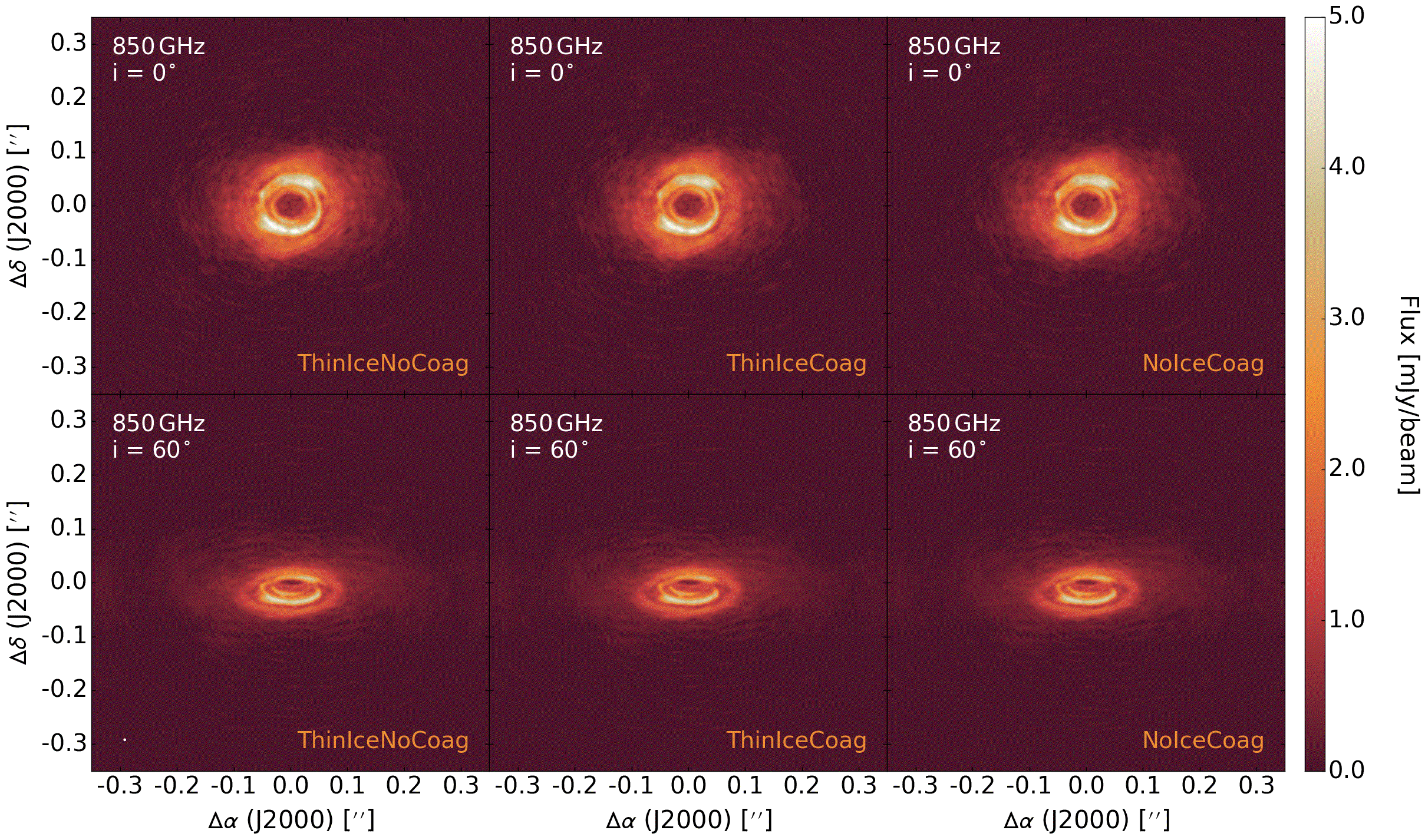}
    \caption{Same as Figure \ref{fig:300GHzFullALMA} but synthesised at 850\,GHz with a beamsize of 0.005$\,\times\,$0.005\,arcsec and a noise level of approximately 25\,\textmu Jy per beam.}
    \label{fig:850GHzFullALMA}
\end{figure*}

Figures \ref{fig:300GHzFullALMA}-\ref{fig:850GHzFullALMA} represent the highest possible image fidelity \alma\ could achieve using our observational setup described in Section \ref{sec:obs}, assuming the maximum baseline of 16.2\,km is available at all frequencies. As can be seen, the spiral structure is very prominent at 300\,GHz, with a S/N \textgreater 30, due to the beamsize reduction at higher frequencies. Unfortunately, the sensitivity is inversely proportional to the frequency, hence at 430\,GHz the spirals are not as easily identified because noise becomes significant. In fact, because of this, the spirals are only distinguishable when assuming a very low dust opacity at 430\,GHz, and at 850\,GHz the spirals are completely dominated by the noise. Note that we do not show the observations at 90\,GHz and 230\,GHz because they are virtually identical to those performed using the \alma\ Cycle 5 antenna configuration as the maximum baseline is already at the fully operational limit for Band 3 and 6; the only difference is the number of antennas (42 for Cycle 5, 50 for fully operational \alma).

\section[radmc-3d]{Comparison with \radmc}
\label{sec:RADMC3D}

In this work we have used \lime\ to produce continuum emission maps that we have then observed synthetically. However, there are numerous radiative transfer codes and it is very important to compare results between them using consistent models in order to understand the strengths and limitations of each. One such alternative to \lime\ is \radmc, which is a Monte-Carlo radiative transfer code used to produce dust continuum emission images and gas line emission images in LTE regions. \radmc\ can also model non-LTE line emission, but currently, only if local modes are adopted. We input our disc model into \radmc\, using our pre-computed hydrodynamic dust temperatures, adopt the `ThinIceCoag' dust opacities and produce a continuum flux image at 300\,GHz to compare to the \lime\ image produced from eight averaged runs using our optimal setup (\nzero, $w = 0.5$, $2 \times 10^5$ points, $z \geq z_{\tau=3}$).

\begin{figure}
    \centering
    \includegraphics[width = 0.425\textwidth]{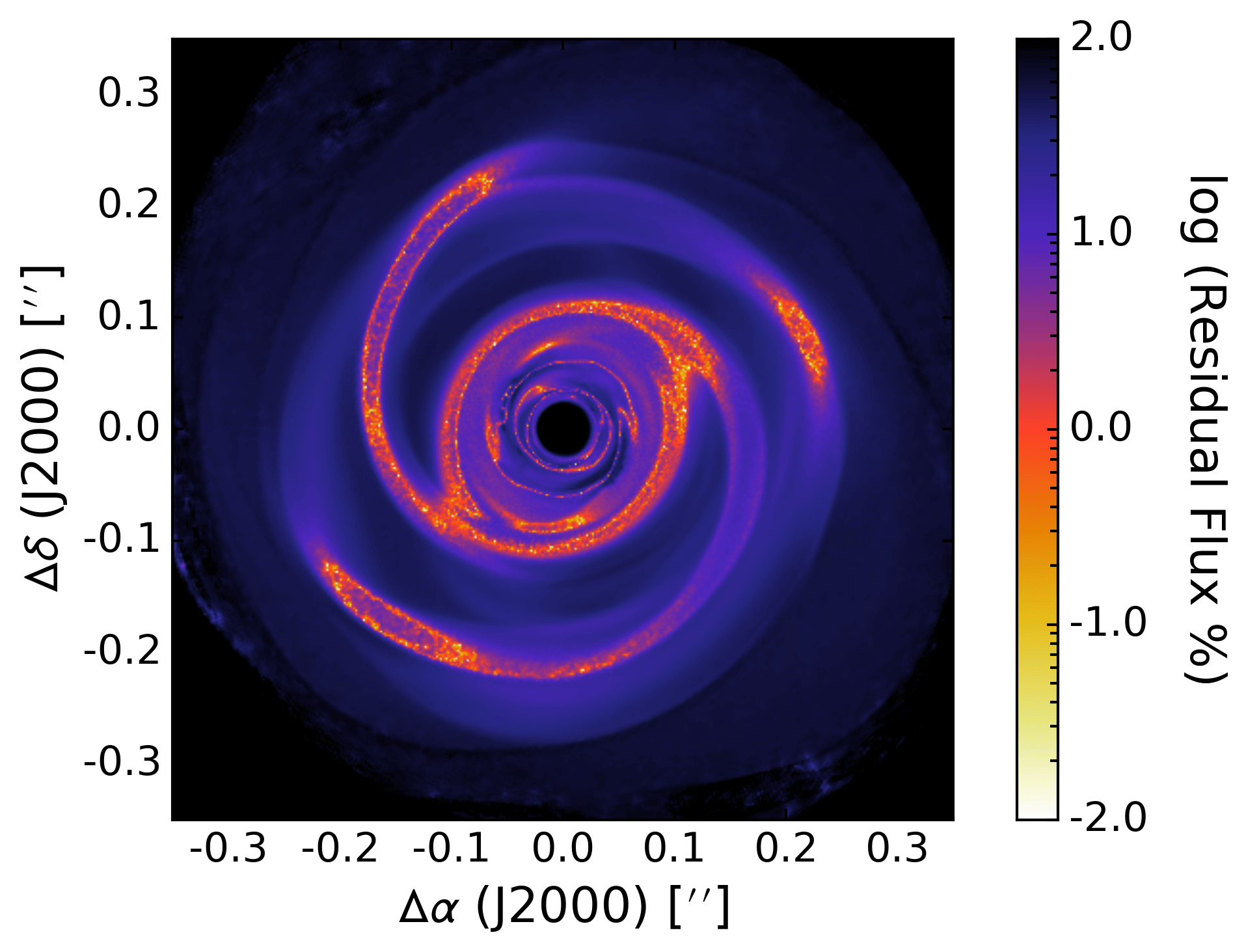}
    \caption{Residual flux between \lime\ runs using $2 \times 10^6$ grid points, $f_L$, and \radmc, $f_R$, using the full hydrodynamic grid. We define this residual as (f$_L$-f$_R$)/(f$_L$+f$_R$), where $f$ denotes flux per pixel.}
    \label{fig:LIMERADMCResidual}
\end{figure}

Figure \ref{fig:LIMERADMCResidual} shows the residual flux between our optimal \lime\ image and the \radmc\ image, and indicates that the two codes produce compatible images as there is a residual difference of \textless 10 per cent within the spiral arms. However there are large discrepancies, up to $\approx$ 60 per cent, in the inter-arm and innermost disc regions that are most likely due to the difference in gridding employed by \lime\ and \radmc. In \radmc, the radiative transfer grid is the entire hydrodynamic model grid, whereas in \lime\ the points are subsampled from this grid and density weighted. Therefore, in less dense regions, and hence optically thinner regions, the grid point spacing in \radmc\ is much finer than in our \lime\ runs. For instance, in the inter-arm regions at the disc midplane, $l$ \textgreater $1.0\,\rm au$, whereas the \radmc\ grid resolution is 0.25\,au; it is important to understand that the mean free path is a statistical average and hence a smaller grid point separation is always preferable if it is achievable. Conversely, in the innermost disc region, where the optical depth is considerably higher, the grid spacing in \lime\ becomes finer than that in \radmc. This may explain why both the inter-arm and innermost disc flux is lower in the \radmc\ continuum emission map, but for opposite reasons.

\begin{figure}
    \centering
    \includegraphics[width = 0.475\textwidth]{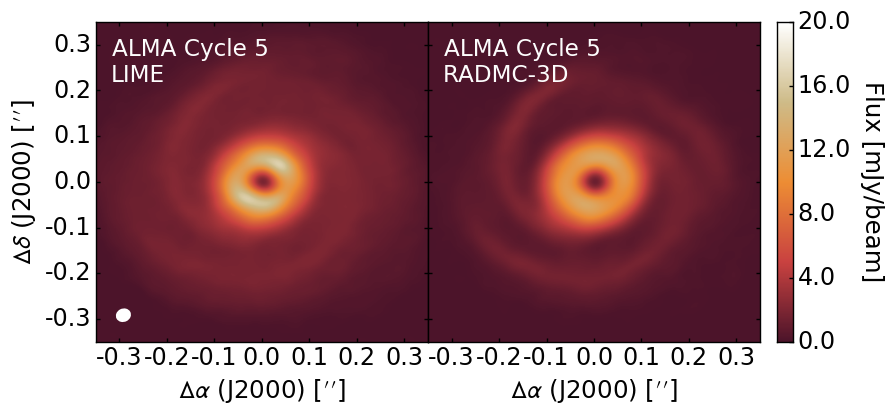}
    \caption{Continuum emission images of \lime\ (left) and \radmc\ (right) radiative transfer calculations of our disc model at 300\,GHz and assuming `ThinIceCoag' dust opacities, synthesised using the \alma\ Cycle 5. The white ellipse in the lower left indicates the size of the beam, which is 0.029\,arcsec and is constant across all panels. The noise is approximately 40\,\textmu Jy per beam across the parameter space.}
    \label{fig:LIMERADMCCycle5}
\end{figure}

\begin{figure}
    \centering
    \includegraphics[width = 0.475\textwidth]{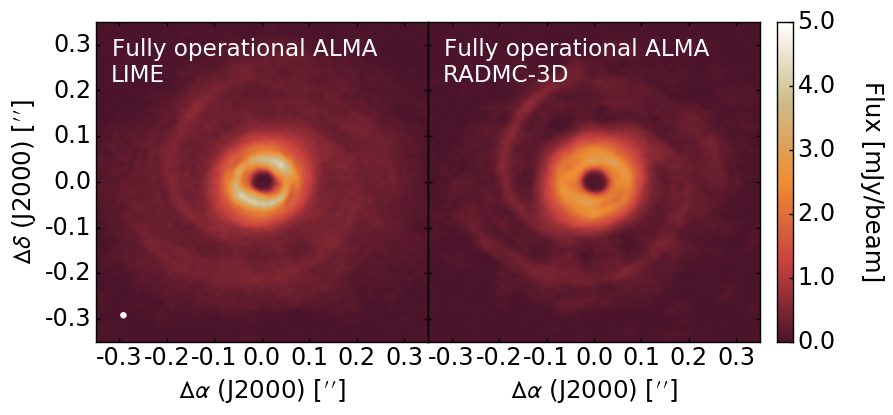}
    \caption{Continuum emission images of \lime\ (left) and \radmc\ (right) radiative transfer calculations of our disc model at 300\,GHz and assuming `ThinIceCoag' dust opacities, synthesised using the \alma\ Cycle 5. The white ellipse in the lower left indicates the size of the beam, which is 0.016\,arcsec and is constant across all panels. The noise is approximately 35\,\textmu Jy per beam across the parameter space.}
    \label{fig:LIMERADMCFullALMA}
\end{figure}

In order to determine if this discrepancy in inter-arm and innermost disc flux leads to observational differences, we produce synthetic observations of our \lime\ and \radmc\ continuum emission maps, using a face-on inclination and the same observational setup as aforementioned. The results for \alma\ Cycle 5 and a fully extended, maximally operational \alma\ are shown in Figures \autoref{fig:LIMERADMCCycle5} and \autoref{fig:LIMERADMCFullALMA} respectively, and demonstrate that, whilst the spiral fluxes are consistent between the two radiative transfer codes as expected, the innermost disc flux is lower in the \radmc\ observations. More significantly for our purposes, however, is that the contrast between arm and inter-arm regions is enhanced in the \radmc\ observations, which results in a more easily detected spiral structure. As this is most likely a result of finer gridding in \radmc\, we conclude that one should perhaps use \radmc\ to produce continuum emission maps when a high resolution model is available.

\smallskip

Whilst \radmc\ likely produces more accurate continuum emission maps than \lime\ for our disc model, at least in the inter-arm regions, we have opted to show results for \lime\ throughout this paper for a number of reasons. Firstly, the continuum emission maps we have produced have only utilised raytracing as the dust temperatures are known \textit{a priori} from the radiative hydrodynamic model. As a result, the computational cost of using a high resolution, regularly spaced grid is low. However, if one were to compute the dust temperatures with a Monte Carlo radiative transfer code, then the computational time would increase significantly as enough photon packages need to be propagated to ensure a sufficient number enter each cell, which becomes especially problematic in the optically thick regions. In this case, using our optimal \lime\ grid may be more computationally efficient than using the high resolution, regularly spaced \radmc\ grid. Secondly, \lime\ is one of the best available tools for producing line emission maps because \lime\ can calculate level populations self-consistently, without the assumption of LTE or optically thin line emission, or the use of the large velocity gradient method that \radmc\ relies upon. Note, however, that whether these assumptions would result in large errors in emission maps is not clear. Nevertheless, when producing line emission maps, the computational time no longer scales linearly with the number of grid points as the level populations for each grid point must be calculated in an iterative sense until convergence is reached. Consequently, it becomes much more important efficiency-wise to optimise the grid sampling, which can be achieved by, for example, restricting the grid points in relation to the continuum optical depth surface (see Section \ref{sec:tausurface}). We plan to explore this avenue of research in a future publication.


\bsp	
\label{lastpage}
\end{document}